\documentclass[11pt,dvipsnames]{article} 
\usepackage{fullpage}

\usepackage{multirow}

\usepackage{xcolor}
\usepackage{url} 
\usepackage{amssymb,amsbsy}
\usepackage{amsmath} 
\usepackage{amsthm}
\usepackage[colorlinks=false]{hyperref}

\usepackage{algorithmic}
\usepackage{algorithm} 
\usepackage{complexity} 
\usepackage{graphicx}
\usepackage{enumerate} 

\newtheorem{theorem}{Theorem}[section]
\newtheorem{theorem*}{Theorem}

\newtheorem{corollary}[theorem]{Corollary}
\newtheorem{lemma}[theorem]{Lemma}

\newtheorem{conjecture}[theorem]{Conjecture}

\newtheorem{definition}[]{Definition}
\newtheorem{claim}[theorem]{Claim}

\newtheorem{fact}[]{Fact}

\newenvironment{myproof}[1]%
{\vspace{1ex}\noindent{\em Pf.}\hspace{0.5em}\def\myproof@name{#1}}%
{\hfill{\tiny \qed\ (\myproof@name)}\vspace{1ex}}

\makeatletter
\newtheorem*{rep@theorem}{\rep@title}
\newcommand{\newreptheorem}[2]{%
\newenvironment{rep#1}[1]{%
 \def\rep@title{#2 \ref{##1} (restated)}%
 \begin{rep@theorem}}%
 {\end{rep@theorem}}}
\makeatother

\newreptheorem{theorem}{Theorem}
\newreptheorem{lemma}{Lemma}
\newreptheorem{conjecture}{Conjecture}
\newreptheorem{proposition}{Proposition}
\newreptheorem{fact}{Fact}

\newenvironment{proof-sketch}{\medskip\noindent{\em Sketch of
Proof.}\hspace*{1em}}{\qed\bigskip}
\newenvironment{proof-attempt}{\medskip\noindent{\em Proof attempt.}\hspace*{1em}}{\bigskip}

\newenvironment{proofof}[1]{\medskip\noindent\emph{Proof of #1.}\hspace*{1em}}{\qed\bigskip}

\newcommand{\inparen }[1]{\left(#1\right)}             
\newcommand{\inbrace }[1]{\left\{#1\right\}}           
\newcommand{\insquare}[1]{\left[#1\right]}             

\newcommand{\abs}[1]{\left|#1\right|}                  

\newcommand{\union}{\cup}
\newcommand{\Union}{\bigcup}
\newcommand{\intersection}{\cap}

\newcommand{\F}{\mathbb{F}}
\newcommand{\N}{\mathbb{N}}

\ifx \@R \@empty

\else

\fi

\ifx \@C \@empty
\newcommand{\C}{\mathbb{C}}
\else
\renewcommand{\C}{\mathbb{C}}
\fi

\renewcommand{\epsilon}{\varepsilon}
\newcommand{\pder}[2]{\partial_{#2}#1}

 \renewcommand{\det}{\text{det}}
\newcommand{\Det}{\text{Det}}

\date{}

\begin{document}

\title{Jacobian hits circuits: Hitting-sets, lower bounds for
  depth-$D$ occur-$k$ formulas \& depth-$3$ transcendence degree-$k$ circuits}

\author{Manindra Agrawal \thanks{Indian Institute of Technology
    Kanpur, {\tt manindra@cse.iitk.ac.in}. Research done while visiting MPI-Informatik under Humboldt Forschungspreise.} 
    \and 
    Chandan Saha  \thanks{Max Planck Institute for Informatics,
    {\tt csaha@mpi-inf.mpg.de}. Supported by IMPECS Fellowship.} 
    \and 
    Ramprasad Saptharishi \thanks{Chennai Mathematical Institute, {\tt ramprasad@cmi.ac.in}. Supported by MSR India Ph.D Fellowship. }
    \and 
    Nitin Saxena \thanks{Hausdorff Center for Mathematics, {\tt ns@hcm.uni-bonn.de}}
    \footnote{The authors are grateful to MPI-Informatik, Saarbr\"{u}cken for its generous hospitality.}
  }

\maketitle

\begin{abstract}
  We present a single, common tool to strictly subsume \emph{all}
  known cases of polynomial time blackbox polynomial identity testing
  (PIT) that have been hitherto solved using diverse tools and
  techniques. In particular, we show that polynomial time hitting-set
  generators for identity testing of the two seemingly different and
  well studied models - depth-$3$ circuits with bounded top fanin,
  and constant-depth constant-read multilinear formulas - can be 
  constructed using one common algebraic-geometry theme: 
  \emph{Jacobian} captures algebraic independence. By exploiting 
  the Jacobian, we design the {\em first} efficient hitting-set generators for broad 
  generalizations of the above-mentioned models, namely:
  \begin{itemize}
  \item depth-$3$ ($\Sigma \Pi \Sigma$) circuits with constant
    \emph{transcendence degree} of the polynomials computed by the
    product gates (\emph{no} bounded top fanin restriction), and
  \item constant-depth constant-\emph{occur} formulas (\emph{no}
    multilinear restriction).
  \end{itemize}
  Constant-\emph{occur} of a variable, as we define it, is a much
  more general concept than constant-read. Also, earlier work on the
  latter model assumed that the formula is multilinear. Thus, our work
  goes further beyond the results obtained by Saxena \&
  Seshadhri (STOC 2011), Saraf \& Volkovich (STOC 2011), Anderson et
  al.\ (CCC 2011), Beecken et al.\ (ICALP 2011) and Grenet et al.\ (FSTTCS 2011), 
  and brings them under one unifying technique.

  In addition, using the same Jacobian based approach, we prove
  exponential lower bounds for the immanant (which includes permanent
  and determinant) on the \emph{same} depth-$3$ and depth-$4$ models
  for which we give efficient PIT algorithms. Our results reinforce the intimate
  connection between identity testing and lower bounds by exhibiting a
  concrete mathematical tool - the Jacobian - that is equally
  effective in solving both the problems on certain interesting and
  previously well-investigated (but not well understood) models of
  computation.

\end{abstract}


\section{Introduction}

\vspace{-0.1in}
A polynomial in many variables, when written down verbosely as a sum
of monomials, might have a humongous expression. \emph{Arithmetic
  circuits}, on the other hand, provide a succinct way to represent
multivariate polynomials. An arithmetic circuit, consisting of
addition ($+$) and multiplication ($\times$) gates, takes several
variables as input and outputs a polynomial in those variables. The
study of arithmetic circuits - as to which algorithmic questions on
polynomials can be resolved efficiently in this model of computation,
and which polynomials do not admit any polynomial-sized circuit
representation - form the foundation of algebraic complexity theory.

One particular algorithmic question, the problem of \emph{polynomial
  identity testing} (PIT), occupies a pivotal position in the theory
of arithmetic circuit complexity. It is the problem of deciding if the
output of a given arithmetic circuit is an identically zero
polynomial. Being such an elementary problem, identity testing has
enjoyed its status of prime importance by appearing in several
fundamental results including primality testing \cite{AKS04}, the
$\PCP$ theorem \cite{ALMSS98} and the $\IP = \PSPACE$ result
\cite{LFKN90,Shamir90}, among many others like graph matching
\cite{L79, MVV87}, polynomial interpolation \cite{CDGK91}, matrix
completion \cite{IKS10}, polynomial solvability \cite{KY08},
factorization \cite{SV-icalp10} and
learning of arithmetic circuits \cite{KS06}. What is more intriguing
is that there is an intimate connection between identity testing and
lower bounds \cite{KI03, HS80, AM10}, especially the problem of
separating the complexity classes $\VP$ from $\VNP$ (which must
necessarily be shown before showing $\P \neq \NP$ \cite{V79,
  SV85}). Proving $\VP \neq \VNP$ amounts to showing that an explicit
class of polynomials, like the Permanent, cannot be represented by
polynomial-sized arithmetic circuits, which in turn would follow if
identity testing can be derandomized using a certain kind of
pseudo-random generator \cite{A05, KI03}. (Note that identity testing
has a simple and efficient randomized algorithm - pick a random point
and evaluate the circuit at it \cite{S80, Z79, DL78}.)

During the past decade, the quest for derandomization of PIT has
yielded several results on restricted models of circuits. But,
fortunately, the search has been made more focussed by a result
\cite{AV08, VSBR83} which states that a polynomial time
\emph{blackbox} derandomization of identity testing for depth-$4$
circuits (via a certain pseudo-random generator) implies a
quasi-polynomial time derandomization of PIT for
\emph{poly-degree}\footnote{Circuits computing polynomials with degree
  bounded by a polynomial function in the size of the circuit.}
circuits. By polynomial time blackbox test, we mean:

\vspace{-0.15in}
\begin{quote}
  \noindent A polynomial time hitting-set generator, which is a
  boolean Turing machine that produces a set of points with
  \emph{small} integer coordinates. These points are then fed (one by
  one) into the circuit, which internally uses \emph{any} arithmetic,
  to output the evaluations of the polynomial at those points. (For
  small characteristic $p$, one works with a field extension, where
  each coordinate of a point is a small vector of integers in
  $\{0,..,p-1\}$.)
\end{quote}
\vspace{-0.15in}
\noindent With depth-$4$ as the final frontier, the results that have
been achieved so far include polynomial time hitting-set generators
for the following models:

\vspace{-0.12in}
\begin{itemize}
\item depth-$2$ ($\Sigma \Pi$) circuits (equivalently, the class of
  {\em sparse} polynomials) \cite{KS01},

  \vspace{-0.12in}
\item depth-$3$ ($\Sigma \Pi \Sigma$) circuits with constant top fanin
  \cite{SS11},

  \vspace{-0.12in}
\item depth-$4$ ($\Sigma\Pi\Sigma\Pi$) multilinear circuits with 
  constant top fanin \cite{SV11},

  \vspace{-0.12in}
\item constant-depth constant-read multilinear formulas \cite{AMV11} 
  (\& sparse-substituted variants),

  \vspace{-0.12in}
\item circuits \emph{generated} by sparse polynomials with constant 
   transcendence degree \cite{BMS11}.
\end{itemize}

\vspace{-0.1in}
\noindent 
To our knowledge, these are the only instances for which polynomial
time hitting-set generators are known. The result on depth-$3$
bounded top fanin circuits is based upon the Chinese Remaindering
technique of \cite{KS07} and the ideal-theoretic framework studied in
\cite{SS10}. Their work followed after a sequence of developments in
rank bound estimates \cite{DS05, KS08, SS09, KS09, SS10}, some using
incidence geometry - although, this result \cite{SS11} in particular
is not rank based. On the other hand, the work on constant-depth
multilinear formulas \cite{AMV11, SV11} is obtained by
building upon and extending the techniques of other earlier results
\cite{KMSV10, SV09, SV08} on `read-once' models. At a high
level, this involved a study of the structure of multilinear formulas under 
the application of partial derivatives with respect to a carefully chosen set 
of variables and invoking depth-$3$ rank bounds (cf. \cite{SY10}
for details). More recently, a third technique has emerged in
\cite{BMS11} which is based upon the concept of \emph{algebraic
  independence} of polynomials. They showed that for any given
poly-degree circuit $C$ and sparse polynomials $f_1,\ldots,f_m$ with
constant transcendence degree, a hitting-set generator for
$C(f_1,\ldots,f_m)$ can be constructed in polynomial time.\\

\vspace{-0.1in}
\noindent \textbf{\large Our contribution - }With these diverse
techniques floating around the study of hitting-set generators, one
wonders: Could there be one single tool that is sufficiently
powerful to capture all these models? Is there any unique feature
underlying these seemingly different models that can lend itself to
the conception of such a unifying tool? The answer to both these
questions, as we show in this work, is \emph{yes}. The key to this
lies in studying the properties of the \emph{Jacobian}, a mathematical
object lying at the very core of algebraic independence. And as for
the `unique feature', notice that in the above four models some
\emph{parameter} of the circuit is \emph{bounded} - be it bounded top
fanin, bounded read of variables, or bounded transcendence
degree. (Bounded depth should not be seen as an extra restriction on
the circuit model because of \cite{AV08}). At an intuitive level, it
seems to us that it is this `bounded parameter'-ness of the circuit
that makes the Jacobian perform at its best.

In the process of finding a universal technique, we strengthen the
earlier results significantly. We construct hitting-set generators not
only for depth-$3$ circuits with bounded top fanin, but also for
circuits of the form $C(T_1, \ldots, T_m)$, where $C$ is a poly-degree
circuit and $T_1, \ldots, T_m$ are products, of linear polynomials, with
bounded transcendence degree. In case of depth-$3$ circuits, $C(T_1,
\ldots, T_m)$ is simply $T_1 + \ldots + T_m$. Further, we remove the
restriction of multilinearity totally from the constant-depth
constant-read model and construct the \emph{first} hitting-set
generator for this class. The condition of constant-read is also
replaced by the more general notion of constant-\emph{occur}.

At this point, one is faced with a natural question: how effective is
this new tool in proving lower bounds? The intimate connection between
efficient algorithms and lower bounds has recurrently appeared in
various contexts \cite{W11, R08, U03, PSZ00, IW97}. For arithmetic
circuits, this link is provably tight \cite{KI03, A05, AV08}:
Derandomizing identity testing is \emph{equivalent} to proving circuit
lower bounds. Which means, one might have to look for techniques that
are powerful enough to handle the dual worlds of algorithm design and
lower bounds with equal effectiveness - for e.g. the \emph{partial
  derivative technique} has been used to prove lower bounds and
identity testing (albeit non-blackbox) on restricted
models (survey \cite{CKW11}); the \emph{$\tau$-conjecture} is another
such example \cite{GKPS11}. In this work, we demonstrate a third tool
- the Jacobian - using which we prove exponential lower bounds for the
immanant (which includes determinant and permanent) on the \emph{same}
depth-$3$ and depth-$4$ models for which we give efficient PIT algorithms. In
particular, this includes depth-$4$ constant-occur formulas, depth-$4$
circuits with constant transcendence degree of the underlying sparse 
polynomials (which significantly generalizes the lower bound result in 
\cite{GKPS11}), and depth-$3$ circuits with constant transcendence 
degree of the polynomials computed by the product gates. To our knowledge, 
all these lower bounds are new and it is not known how to prove them
using earlier techniques. (A gist of this paper is provided in 
Figure \ref{fig-1}, Section \ref{sec-conclusion}.)


\vspace{-0.1in}
\subsection{A tale of two PITs (\& three lower bounds)}

\vspace{-0.06in} A set of polynomials $ \mathbf{f} =
\inbrace{f_1,\cdots, f_m} \subset \F[x_1,\cdots, x_n]$ (in short,
$\F[\mathbf{x}]$) is \emph{algebraically independent} over $\F$ if
there is no nonzero polynomial $H\in \F[y_1,\cdots, y_m]$ such that
$H(f_1,\cdots, f_m)$ is identically zero. A maximal subset of
$\mathbf{f}$ that is algebraically independent is a
\emph{transcendence basis} of $\mathbf{f}$ and the size of such a
basis is the \emph{transcendence degree}\footnote{Since algebraic
  independence satisfies the matroid property cf. \cite{O92},
  transcendence degree is well-defined.} of $\mathbf{f}$ (denoted
$\mathrm{trdeg}_\F \hspace{0.025in} \mathbf{f}$). Our first theorem
states:

\vspace{-0.05in}
\begin{theorem} \label{thm:gend3-bndtrdegPIT} Let $C$ be a poly-degree
  circuit of size $s$ and each of $T_1, \ldots, T_m$ be a product of
  $d$ linear polynomials in $\F[x_1, \ldots, x_n]$ such that
  $\mathrm{trdeg}_{\F} \hspace{0.025in} \{T_1, \ldots, T_m\} \leq r$. A
  hitting-set for $C(T_1, \ldots, T_m)$ can be constructed
  in time polynomial in $n$ and $(sd)^r$, assuming $\emph{char}(\F) =
  0$ or $ > d^r$.
\end{theorem}

\vspace{-0.05in}
\noindent If $C$ is a single $+$ gate, we get a hitting-set generator
for depth-$3$ circuits with constant \emph{transcendence degree} of the
polynomials computed by the product gates (there is \emph{no}
restriction on top fanin).

Our second result uses the following generalization of \emph{read-$k$}
formulas (where every variable appears in at most $k$ leaf nodes of
the formula) to \emph{occur-$k$} formulas. Two reasons behind this
generalization are: One, to accommodate the power of exponentiation -
if we take the $e$-th power of a read-$k$ formula using a product
gate, the `read' of the resulting formula goes up to $ek$ - we would
like to avoid this superfluous blow up in read. Two, a read-$k$
formula has size $O(kn)$, which severely hinders its power of
computation - for instance, determinant and permanent cannot even be
expressed in this model when $k$ is a constant \cite{K85}. This calls
for the following definition.

\vspace{-0.05in}
\begin{definition} \label{defn:occurk} An {\em occur-$k$ formula} is a
  rooted tree with internal gates labelled by $+$ and $\times\curlywedge$ 
  (power-product gate). A $\times\curlywedge$ gate, on inputs $g_1,\ldots,g_m$ 
  with incoming  edges labelled $e_1,\ldots,e_m\in\N$, computes $g_1^{e_1}
  \cdots g_m^{e_m}$. At the leaves of
  this tree are depth-$2$ formulas computing sparse polynomials ({\em leaf nodes}), where
  every variable occurs in at most $k$ of these sparse polynomials.
\end{definition}

\vspace{-0.05in}
\noindent Size of a $\times\curlywedge$ gate is defined as the integer 
$(e_1+\cdots+e_m)$ associated with its incoming edges, while size of a $+$ 
gate is counted as one. Size of a leaf node is the size of the
corresponding depth-$2$ formula. With these conventions, \emph{size}
of an occur-$k$ formula is defined to be the total size of all its gates (and
leaf nodes) plus the number of edges. 
\emph{Depth} is defined to be the number of
layers of $+$ and $\times\curlywedge$ gates plus $2$ (the `plus $2$' accounts for
the depth-$2$ formulas at the leaves). Thus, occur-$k$ is more relaxed than
the traditional read-$k$ as it packs the ``power of powering'' (to borrow 
from \cite{GKPS11}), and the leaves are sparse polynomials (at most $kn$
many) whose dependence on its variables is arbitrary. E.g. 
$(x_1^3x_2+x_1^2x_3^2+x_1x_4)^e$ is {\em not} read-$1$ but is trivially 
depth-$3$ occur-$1$.

\vspace{-0.05in}
\begin{theorem} \label{thm:dDkrPIT} A hitting-set for a
  depth-$D$ occur-$k$ formula of size $s$ can be constructed in time
  polynomial in $s^R$, where $R =(2k)^{2D\cdot 2^D}$ (assuming
  $\emph{char}(\F) = 0$ or $> s^R$).
\end{theorem}

\vspace{-0.05in} A tighter analysis for depth-$4$ occur-$k$ formulas
yields a better time complexity. 
Note that a depth-$4$ occur-$k$ formula allows unbounded top fanin. Also, it 
can be easily seen to subsume $\Sigma\Pi\Sigma\Pi(k)$ multilinear circuits 
studied by \cite{SV11, KMSV10}.

\vspace{-0.05in}
\begin{theorem} \label{thm:d4krPIT} A hitting-set for a
  depth-$4$ occur-$k$ formula of size $s$ can be constructed in time
  polynomial in $s^{k^2}$ (assuming $\emph{char}(\F) = 0$ or
  $> s^{2k}$).
\end{theorem}

\vspace{-0.05in} For constant-depth, the above theorems not only remove 
the restriction of multilinearity (and relax read-$k$ to occur-$k$), but further improve
upon the time complexity of \cite{AMV11} and \cite{SV11}. The hitting-set 
generator of \cite{AMV11} works in time $n^{k^{O(k^2)} + O(kD)}$,
which is super-exponential when $k = \Omega(s^{\epsilon/2D \cdot 2^D})$ for
any positive $\epsilon < 1$ and a constant $D$, whereas the generator
in Theorem \ref{thm:dDkrPIT} runs in sub-exponential time for the same
choice of parameters. The running time of \cite{SV11} is $s^{O(k^3)}$, which is 
slightly worse than that of Theorem \ref{thm:d4krPIT}.

Since any polynomial has an exponential-sized depth-$2$, occur-$1$
formula (just the sparse representation), proving lower bounds on this
model is an interesting proposition in its own right.

\vspace{-0.03in}
\begin{definition} \label{defn:immanant} \cite{LR34} For any character
  $\chi:S_n\rightarrow \C^{\times}$, the {\em immanant} of a matrix $M =
  (x_{ij})_{n \times n}$ with respect to $\chi$ is defined as
  $\text{Imm}_\chi(M) = \sum_{\sigma\in S_n}
  \chi(\sigma)\prod_{i=1}^{n}{ x_{i\sigma(i)}}$.
\end{definition}

\vspace{-0.03in}
\noindent Determinant \& permanent are special cases of the immanant
with $\chi$ as the alternating sign character \& the identity
character, respectively. Denote $\text{Imm}_\chi(M)$ by $\text{Imm}_n$
for an arbitrarily fixed $\chi$.

\vspace{-0.06in}
\begin{theorem}\label{thm:lowerbnd-depth4-occurk}
  Any depth-$4$ occur-$k$ formula that computes $\text{Imm}_n$ must
  have size $s = 2^{\Omega\inparen{{n/k^2}}}$ over any field of
  characteristic zero (even counting each $\times\curlywedge$ gate as size one).
\end{theorem}

\vspace{-0.05in}
\noindent Thus, if each variable occurs in at most $n^{1/2 - \epsilon}$ 
$(0 < \epsilon < 1/2)$ many underlying sparse polynomials, it takes an exponential
sized depth-$4$ circuit to compute $\text{Imm}_n$. Our next result is
an exponential lower bound on the model for which hitting-set was
developed in \cite{BMS11} (but no lower bound was shown). It is also an
improvement over the result obtained in \cite{GKPS11} which
holds only for more restricted depth-$4$ circuits over reals.

\vspace{-0.05in}
\begin{theorem}\label{thm:lowerbnd-sparse-algRank}
  Let $C$ be any circuit. Let $f_1, \ldots, f_m$ be sparse polynomials
  (of any degree) with sparsity bounded by $s$ and their $\mathrm{trdeg}$ 
  bounded by $r$. If $C(f_1, \ldots, f_m)$ computes $\text{Imm}_n$ then $s =
  2^{\Omega(n/r)}$ over any field of characteristic zero.
\end{theorem}

\vspace{-0.05in}
\noindent Which means, any circuit involving fewer than $n^{1 -
  \epsilon}$ $\Sigma\Pi$-polynomials at the last levels, must have
exponential size to compute $\text{Imm}_n$. (The models of
Theorem \ref{thm:lowerbnd-depth4-occurk} $\&$
\ref{thm:lowerbnd-sparse-algRank} are incomparable). The next result
is on the model for which hitting-set is given by Theorem
\ref{thm:gend3-bndtrdegPIT}.

\vspace{-0.03in}
\begin{theorem} \label{thm:lowerbound-depth3} Let $C$ be any circuit
  and $T_1, \ldots, T_m$ be products of linear polynomials. If $C(T_1,
  \ldots,$ $T_m)$ computes $\text{Imm}_n$ then $\mathrm{trdeg}_{\F}
  \hspace{0.025in} \{T_1, \ldots, T_m\} = \Omega(n)$ over any field of
  characteristic zero.
\end{theorem}

\vspace{-0.05in}
\noindent Which means, any circuit involving $o(n)$ 
$\Pi\Sigma$-polynomials at the last levels {\em cannot} compute $\text{Imm}_n$.

\vspace{-0.1in}
\subsection{Our ideas}

\vspace{-0.05in} The exact reasons why our techniques work, where older ones failed, 
are extremely technical. However, we now give the motivating, but imprecise, ideas.
To a set of products of sparse polynomials $\{T_1,\ldots,T_m\}$ we associate a polynomial --
the Jacobian $J(T_1,\ldots,T_r)$. It captures the algebraic independence of 
$T_1,\ldots,T_r$ (assuming this to be a transcendence basis of the $T_i$'s). If we 
could find an $r$-variate linear map $\varphi$ that keeps $\varphi\circ J(T_1,\ldots,T_r)$ nonzero, 
then $\varphi(T_1),\ldots,\varphi(T_r)$ are again algebraically independent and it can be shown
that for {\em any} $C$: $C(T_1,\ldots,T_m)=0$ iff  $C(\varphi(T_1),\ldots,\varphi(T_m))=0$. 
Since $T_i$'s are not sparse, the Jacobian is usually a difficult polynomial to work with, and 
so is finding $\varphi$. However, for the special models in this paper we are able to design 
$\varphi$ - mainly because Jacobian (being defined via partial derivatives) has a nice `linearizing
effect', on the circuit product gates, that factors itself. The $\varphi$ ultimately provides a 
hitting-set for 
$C(T_1,\ldots,T_m)$, as we reduce to a situation where $r$ is constant. 

The initial idea for lower bounds is similar. Suppose $\text{Imm}_n=$ $C(T_1,\ldots, T_m)$. 
Then, by algebraic dependence, $J(\text{Imm}_n,T_1,\ldots, T_r)=0$. Our proofs then exploit 
the nature of this identity for the special models. This part requires proving several combinatorial 
properties of the immanant.

\vspace{-0.15in}
\section{Preliminaries: Jacobian and faithful homomorphisms}
\vspace{-0.1in} Our contribution, in this section, is an elementary
proof of Theorem \ref{thm:faithful-pit}, which was originally proved
in \cite{BMS11} using Krull's {\em Hauptidealsatz}. Here, we state the main
properties of the Jacobian and faithful homomorphisms without proofs -
for details, refer to \cite{BMS11b} (or Appendix \ref{sec:app_prel}).
\begin{definition}
  The \emph{Jacobian} of polynomials $\mathbf{f} = \inbrace{f_1,
    \cdots, f_m}$ in $\F[x_1,\cdots, x_n]$ is the matrix
  $\mathcal{J}_{\mathbf{x}}(\mathbf{f}) = (\pder{f_i}{x_j})_{m \times
    n}$, where $\pder{f_i}{x_j} = \partial f_i/ \partial x_j$. Let $S
  \subseteq \mathbf{x} = \{ x_1, \ldots, x_n\}$ and $\abs{S} =
  m$. Then $J_{S}(\mathbf{f})$ denote the minor of
  $\mathcal{J}_{\mathbf{x}}(\mathbf{f})$ formed by the columns
  corresponding to the variables in $S$.
\end{definition}
\begin{fact}[Jacobian criterion] \label{fact:jacobi-criterion} Let $\mathbf{f} \subset
  \F[\mathbf{x}]$ be a finite set of polynomials of degree at most
  $d$, and $\mathrm{trdeg}_{\F} \hspace{0.025in} \mathbf{f} \leq r$. If
  $\emph{char}{(\F)} = 0$ or $\emph{char}{(\F)} > d^r$, then
  $\mathrm{trdeg}_{\F} \hspace{0.025in} \mathbf{f}=$ 
  $\mathrm{rank}_{\F(\mathbf{x})} \mathcal{J}_{\mathbf{x}}(\mathbf{f})$.
\end{fact}
\begin{fact}[Chain rule]\label{fact:jacobian-chain-rule}
  For any finite set of polynomials $\mathbf{f} \subset
  \F[\mathbf{x}]$ and a homomorphism $\Phi:\F[\mathbf{x}] \rightarrow
  \F[\mathbf{y}]$, we have $\mathcal{J}_{\mathbf{y}}(\Phi(\mathbf{f}))
  = \Phi\inparen{\mathcal{J}_{\mathbf{x}}(\mathbf{f})} \cdot
  \mathcal{J}_{\mathbf{y}}(\Phi(\mathbf{x}))$.
\end{fact}

\begin{definition}
  A homomorphism $\Phi: \F[\mathbf{x}] \rightarrow \F[\mathbf{y}]$
  ($\mathbf{y}$ is another set of variables) is said to be
  \emph{faithful} to a finite set of polynomials $\mathbf{f} \subset
  \F[\mathbf{x}]$ if $\mathrm{trdeg}_{\F} \hspace{0.025in} \mathbf{f} =
  \mathrm{trdeg}_{\F} \hspace{0.025in} \Phi(\mathbf{f})$.
\end{definition}
\begin{theorem}[Faithful is useful] \label{thm:faithful-pit} Let $\mathbf{f} =
  \inbrace{f_1,\cdots, f_m} \subset \F[\mathbf{x}]$ and $\Phi$ be a
  homomorphism faithful to $\mathbf{f}$. For any polynomial
  $C\in\F[y_1,\cdots, y_m]$, $C(\mathbf{f}) = 0
  \Leftrightarrow C(\Phi(\mathbf{f})) = 0$.
\end{theorem}
\begin{lemma}[Vandermonde is faithful]\label{lem:composition-lemma}
  Let $\mathbf{f} \subset \F[\mathbf{x}]$ be a finite set of
  polynomials of degree at most $d$, 
  $\mathrm{trdeg}_{\F} \hspace{0.025in} \mathbf{f} \leq r$, and
  $\emph{char}{(\F)} = 0$ or $> d^r$. Let 
  $\Psi:\F[\mathbf{x}]\rightarrow \F[\mathbf{z}]$ be a homomorphism such that
  $\mathrm{rank}_{\F(\mathbf{x})} \mathcal{J}_{\mathbf{x}}(\mathbf{f})$ $=$
  $\mathrm{rank}_{\F(\mathbf{z})} \Psi( \mathcal{J}_{\mathbf{x}}(\mathbf{f}) )$.
  
  Then, the map $\Phi:\F[\mathbf{x}]\rightarrow$ 
  $\F[\mathbf{z},t, y_1, \ldots, y_r]$ that maps, for all $i$, 
  $x_i \mapsto \inparen{\sum_{j=1}^r y_j t^{ij}} + \Psi(x_i)$ is 
  a homomorphism faithful to $\mathbf{f}$.
\end{lemma}
\noindent The proof of the above lemma is based upon Facts 
\ref{fact:jacobi-criterion}, \ref{fact:jacobian-chain-rule} and an 
application of `rank preserving' linear maps \cite{GR05}. 
See Appendix \ref{sec:app_prel}.

\vspace{-0.1in}
\section{Hitting-set for constant transcendence degree depth-$3$
  circuits}\label{sec:PIT-d3}

\vspace{-0.1in} Let $C$ be \emph{any} circuit and $D$ be the circuit
$C(T_1, \cdots, T_m)$, where each $T_i$ is of the form $\prod_{j=1}^d
\ell_{ij}$, every $\ell_{ij}$ is a linear polynomial in $\F[x_1, \ldots,
x_n]$. Denote by $\mathbf{T}$ the set $\{ T_1, \ldots, T_m\}$ and by
$L(T_i)$ the multiset of linear polynomials that constitute
$T_i$. Suppose $\mathrm{trdeg}_{\F} \hspace{0.025in} \mathbf{T} = k \leq
r$ and $\mathbf{T}_k = \{T_1, \ldots, T_k\}$ be a transcendence basis
of $\mathbf{T}$. Since $\mathcal{J}_{\mathbf{x}}(\mathbf{T}_k)$ has
full rank ($\mathrm{char}(\F) = 0$ or $\mathrm{char}(\F) > d^r$), without
loss of generality assume that the columns corresponding to
$\mathbf{x}_k = \{x_1,\cdots, x_k\}$ form a nonzero $k \times k$
minor of $\mathcal{J}_{\mathbf{x}}(\mathbf{T}_k)$. By
Lemma~\ref{lem:composition-lemma}, if we construct a $\Psi:
\F[\mathbf{x}] \rightarrow \F[\mathbf{z}]$ that keeps
$J_{\mathbf{x}_k}(\mathbf{T}_k)$ nonzero then $\Psi$ can easily be
extended to a homomorphism $\Phi: \F[\mathbf{x}] \rightarrow 
\F[\mathbf{z},t, y_1,\ldots, y_r]$ that is faithful to $\mathbf{T}$. And
hence, by Theorem \ref{thm:faithful-pit}, it would follow that
$\Phi(D) = 0$ if and only if $D = 0$.

If $T_i = \prod_{j=1}^{d}{\ell_{ij}}$ then $\pder{T_i}{x} = T_i \cdot
\inparen{\sum_{j=1}^{d}{\pder{\ell_{ij}}{x}/\ell_{ij}}}$. By expanding,
using this additive structure of $\pder{T_i}{x}$ and the linearity of determinant
wrt rows, the determinant
$J_{\mathbf{x}_k}(\mathbf{T}_k)$ takes the following form,

\vspace{-0.2in}
\begin{equation} \label{eqn:jacobiansum}
  J_{\mathbf{x}_k}(\mathbf{T}_k) = \sum_{\ell_1 \in L(T_1), \ldots,
    \ell_k \in L(T_k)} \frac{T_1\cdots T_k}{\ell_1\cdots \ell_k}\cdot
  J_{\mathbf{x}_k}(\ell_1,\cdots, \ell_k).
\end{equation}

\vspace{-0.05in}
\noindent Call a set of linear polynomials \emph{independent} if the
correponding homogenous linear parts (i.e. the constant-free parts)
are $\F$-linearly independent. The term
$J_{\mathbf{x}_k}(\ell_1,\cdots, \ell_k)$ ensures that the above sum
is only over those $\ell_1,\cdots, \ell_k$ that are independent linear
polynomials (otherwise the Jacobian vanishes). The sum has the form of
a depth-$3$ circuit, call it $H_0$, and we intend to construct a low-variate
$\Psi$ such that $\Psi(H_0) \neq 0$. We show that this is achieved by
a $\Psi$ that preserves the independence of a `small' set of linear
polynomials - which we call a \emph{certificate} of $H_0$.

\vspace{0.06in}
\noindent \textbf{Certificate of $H_0$:} We can assume that the terms
$J_{\mathbf{x}_k}(\ell_1,\cdots, \ell_k)$ in equation
(\ref{eqn:jacobiansum}) are \emph{nonzero} field constants. Let
$\mathcal{L}(H_0)$ be the set of all linear polynomials occurring in
the denominator terms ``$\ell_1\cdots \ell_k$'' of all the summands in
sum (\ref{eqn:jacobiansum}). By adjusting the field constants
at the numerators, we can
assume that no two linear polynomials in $\mathcal{L}(H_0)$ are
constant multiple of each other. This means, the depth-$3$ circuit
$H_0$ has the form $H_0 = T \cdot \sum_{L}{\alpha_L/\ell_1 \cdots
  \ell_k}$, where $T := \prod_{i=1}^{k}{T_k}$, $\alpha_L$ is a nonzero
field constant and the sum runs over \emph{some} sets $L = \{
\ell_1,\cdots, \ell_k\}$ of $k$ independent linear polynomials in
$\mathcal{L}(H_0)$. Define, \emph{content} of a depth-$3$ circuit $G =
\sum_{i}{P_i}$, where $P_i$ is a product of linear polynomials, as
$\text{cont}(G) := \gcd_i\{P_i\}$, and let the {\em simple part} 
$\text{sim}(G) := G/\text{cont}(G)$. Hence 
$\text{cont}(H_0) = \gcd_L\{T/\ell_1 \cdots\ell_k\}$ and

\vspace{-0.15in}
\begin{equation}
  \text{sim}(H_0) = F_0 \cdot \sum_{L}{\frac{\alpha_L}{\ell_1 \cdots \ell_k}}, \text{ where $F_0 = \frac{T}{\text{cont}(H_0)}$},
\end{equation}

\vspace{-0.05in} Note that, since $\ell\in\mathcal{L}(H_0)$ iff $\ell||F_0$, 
$F_0$ is simply the product of the linear
polynomials in $\mathcal{L}(H_0)$ and so $\deg(F_0) =
\abs{\mathcal{L}(H_0)}$.  For any $\ell \in \mathcal{L}(H_0)$, the
terms in $\text{sim}(H_0)$ that survive modulo $\ell$ are those with
$\ell$ in the denominator ``$\ell_1 \cdots \ell_k$'' of the above
expression. Hence, $ H_1 := \text{sim}(H_0) \bmod \ell_1 = F_0/\ell_1
\cdot \sum_{\ell_2,\cdots, \ell_k} {\alpha_L/\ell_2\cdots \ell_k}$.
We can treat $H_1$ as a depth-$3$ circuit in one less variable:
Suppose that $\ell_1 = c_1x_1 + \sum_{i=2}^{n}{c_ix_i}$ where 
$c_i\text{'s} \in\F$ and $c_1 \neq 0$, then we can replace $x_1$ by
$-\sum_{i=2}^{n}{c_ix_i}/c_1$ in $\text{sim}(H_0)$, particularly in
$F_0/\ell_1$ (of course, after dividing $F_0$ by $\ell_1$) as well as
in each of $\ell_2, \ldots, \ell_k$ in the denominators, so that $H_1$
becomes a depth-$3$ circuit in $\F[x_2, \ldots, x_n]$. Therefore, it
makes perfect sense to talk about $\text{cont}(H_1)$ and
$\text{sim}(H_1)$. Observe that $\ell_2,\cdots, \ell_k$ remain
independent linear polynomials modulo $\ell_1$, and so $H_1$ is a
depth-$3$ circuit of the `same nature' as $H_0$ but with one less
linear polynomials in the denominators. Also, the linear polynomials
in $\mathcal{L}(H_1)$ is a subset of the linear polynomials in
$\mathcal{L}(H_0)$ modulo $\ell_1$. Extending the above argument, it
is possible to define a sequence of circuits: $H_i :=
\text{sim}(H_{i-1}) \mod \tilde{\ell}_i$, ($1 \leq i \leq k$) where
$\tilde{\ell_i} \in \mathcal{L}(H_{i-1})$. Further, $\mathcal{L}(H_i)$
is a subset of $\mathcal{L}(H_{i-1})$ modulo $\tilde{\ell_i}$, which
implies that essentially there are independent linear polynomials, say
$\ell_1, \ldots, \ell_k$, in $\mathcal{L}(H_0)$ such that
$\tilde{\ell_i} = \ell_i \bmod (\ell_1, \ldots, \ell_{i-1})$ and
therefore $H_i = \text{sim}(H_{i-1}) \mod (\ell_1, \ldots, \ell_i)$.

\vspace{-0.05in}
\begin{lemma}[Certifying path]\label{lem:path-cert-rational-sum}
  There exists independent linear polynomials $\inbrace{\ell_1,\cdots,
    \ell_k} \subseteq \mathcal{L}(H_0)$ such that $H_i \neq 0 \bmod
  (\ell_1, \ldots, \ell_i)$, $\forall i\in[k]$, and $H_k$ is a nonzero product of linear
  polynomials in $\mathcal{L}(H_0)$ modulo $(\ell_1,\cdots, \ell_k)$. \\
  {\em Proof: Induction on $k$, see Appendix \ref{sec:app_hs-d3}.}
\end{lemma}


\vspace{-0.05in}
\noindent A set $\{\ell_1, \ldots, \ell_k \}$, satisfying Lemma
\ref{lem:path-cert-rational-sum}, is called a \emph{certifying path} of
$H_0$. Fix a certifying path $\inbrace{\ell_1,\cdots, \ell_k}$. Let $\Psi:
\F[\mathbf{x}] \rightarrow \F[z_1, \ldots, z_{k+1}]$ be such that
$\Psi(\ell_1), \ldots, \Psi(\ell_k)$ are independent linear
polynomials in $\F[\mathbf{z}]$ and for every $\ell \in
\cup_{i=1}^{k}{L(T_i)}$, $\ell \neq0 \bmod \inparen{\ell_1, \ldots, \ell_k}$ 
iff $\Psi(\ell) \neq 0\bmod \inparen{\Psi(\ell_1), \ldots, \Psi(\ell_k)}$. We call such a
$\Psi$ a \emph{rank-$(k+1)$ preserving map} for $H_0$. It can be shown
that one of the maps $\Psi_b: x_i \mapsto
\sum_{j=1}^{k+1}{z_jb^{ij}}$, where $b$ runs over $dkn(k+1)^2$ distinct
elements of $\F$, is a rank-$(k+1)$ preserving map for $H_0$. (It is a
simple application of \cite{GR05}. See Corollary~\ref{cor:gabizon-raz}). 

\vspace{-0.05in}
\begin{theorem}[Certificate] \label{thm:varred} If $\Psi:\F[\mathbf{x}] \rightarrow
  \F[z_1,...,z_{k+1}]$ is a rank-$(k+1)$ preserving map for $H_0$, then
  $\Psi(H_0) \neq 0$.\\
  {\em Proof: Reverse induction on $k$, see Appendix \ref{sec:app_hs-d3}.}
\end{theorem}


\vspace{-0.08in}
\begin{proofof}{Theorem~\ref{thm:gend3-bndtrdegPIT}}
As $r\geq k$, we can assume that the rank-$(k+1)$ preserving map $\Psi$ is in fact
a map from $\F[\mathbf{x}] \rightarrow \F[z_1, \ldots, z_{r+1}]$. 
Therefore, by Lemma \ref{lem:composition-lemma}, $\Phi$ is a map
from $\F[\mathbf{x}] \rightarrow \F[y_1, \ldots y_r, t, z_1, \ldots,
z_{r+1}]$ such that: $D=0$ iff $\Phi(D) = 0$. Since $C$ is a
poly-degree circuit of size $s$, $\Phi(C(T_1, \ldots, T_m))$ is a
polynomial of degree at most $ds^{O(1)}$ resp. $nrds^{O(1)}$ in the variables
$\mathbf{y}, \mathbf{z}$ resp. $t$. Using \cite{S80, Z79, DL78} lemma, we can 
construct a hitting-set
for $\Phi(D)$ in time polynomial in $n(sd)^r$. Since construction of
$\Psi$ takes time $\poly(ndr)$, the total time taken is
$\poly(n,(sd)^r)$.
\end{proofof}

\vspace{-0.25in}
\section{Hitting-set for constant-depth constant-occur
  formulas} \label{sec:depthDoccurkPIT}

\vspace{-0.06in} \textbf{Bounding the top fanin -} Let $C$ belong to
the class $\mathcal{C}$ of depth-$D$ occur-$k$ formulas of size
$s$. Observe that if $C(x_1, \ldots, x_n)$ is non-constant and nonzero, then 
there is an $i$ such that $\tilde{C} := C(x_1,\cdots,
x_{i-1},x_i+1,x_{i+1},\cdots, x_n) - C(x_1,\cdots, x_n) \neq 0$,
assuming $\mathrm{char}(\F) > s^D$ (i.e. the bound on the degree of $C$). 
If $C$ has a $+$ gate on top then
$C(\mathbf{x}) = \sum_{i=1}^{m}{T_i}$, where $T_i$'s are computed by
$\times\curlywedge$ gates at the next level. Since $x_i$
occurs in at most $k$ of the $T_i$'s, $\tilde{C}$ has top fanin at
most $2k$. If $C$ has a $\times\curlywedge$ gate on top then
$\tilde{C}$ has a $+$ gate on top with fanin $2$ and
$\text{depth}(\tilde{C}) = D+1$. Therefore, $\tilde{C}$ belongs to the
class $\tilde{\mathcal{C}}$ of depth-$(D+1)$ occur-$2k$ formulas of
size at most $(s^2+s)$, and a $+$ gate on top with fanin bounded by
$2k$. Suppose $\tilde{\mathcal{H}}$ is a hitting-set for the class
$\tilde{\mathcal{C}}$. Form a new set $\mathcal{H} \supset
\tilde{\mathcal{H}}$ by including points $(\alpha_1+1, \alpha_2,
\ldots, \alpha_n), (\alpha_1, \alpha_2+1, \ldots, \alpha_n), \ldots,
(\alpha_1, \ldots, \alpha_{n-1}, \alpha_n+1)$ in $\mathcal{H}$ for
every $(\alpha_1, \alpha_2, \ldots, \alpha_n) \in
\tilde{\mathcal{H}}$. Observe that $\mathcal{H}$ is a hitting-set for
$\mathcal{C}$ and $\text{size}(\mathcal{H}) = n \cdot
\text{size}(\tilde{\mathcal{H}})$. Therefore, it is sufficient if we
construct $\tilde{\mathcal{H}}$. By reusing symbols, assume that $C$
is a depth-$D$ occur-$k$ formula of size $s$ with a $+$ gate on top
having top fanin at most $k$.

Let $C(\mathbf{x}) = \sum_{i=1}^{k}{T_i}$. The goal is to construct a
$\Phi$ that is faithful to $\mathbf{T} = \{T_1, \ldots, T_k\}$. 
Let $\mathbf{T}_r = \{T_1, \ldots,
T_r\}$ be a transcendence basis of $\mathbf{T}$. Since
$\mathcal{J}_{\mathbf{x}}(\mathbf{T}_r)$ has full rank
($\mathrm{char}(\F) = 0$ or $> s^{Dr}$), assume that the columns
corresponding to $\mathbf{x}_r = \{x_1, \ldots, x_r\}$ form a nonzero
minor of $\mathcal{J}_{\mathbf{x}}(\mathbf{T}_r)$. By Lemma
\ref{lem:composition-lemma}, it suffices to construct a $\Psi$ that
keeps $J_{\mathbf{x}_r}(\mathbf{T}_r) \neq 0$.\\

\vspace{-0.1in}
\noindent \textbf{Proof idea -} Identify a gate with the polynomial it
computes, and count \emph{level} of a gate from the top - the gates
$T_i$'s are at level $2$. Suppose each $T_i$ is a $\times\curlywedge$ gate and
$T_i = \prod_{\ell = 1}^{d}{P_{i, \ell}^{e_{i,\ell}}}$, where $P_{i, \ell}$'s are
gates at level $3$. Since $T_i$ is also an occur-$k$ formula, $x_1,
\ldots, x_r$ appear in at most $kr$ of the $P_{i, \ell}$'s, say $P_{i,1}, 
\ldots P_{i,kr}$. Hence, $\partial_j T_i =
(\prod_{\ell = kr+1}^{d}{P_{i, \ell}^{e_{i,\ell}} }) \cdot (\partial_j
\prod_{\ell=1}^{kr}{P_{i, \ell}^{e_{i,\ell}} })$ for every $1 \leq i,j \leq r$ and
therefore, $J_{\mathbf{x}_r}(\mathbf{T}_r) =
(\prod_{i=1}^{r}{\prod_{\ell = kr + 1}^{d}{P_{i, \ell}^{e_{i,\ell}} }}) \cdot
\det(\partial_j \prod_{\ell=1}^{kr}{P_{i, \ell}^{e_{i,\ell}} })$. Now notice that
$\det(\partial_j \prod_{\ell=1}^{kr}{P_{i, \ell}^{e_{i,\ell}} })$ is a polynomial in
$P_{i, \ell}$ and $\partial_j P_{i, \ell}$, for $1 \leq i, j \leq r$ and
$1 \leq \ell \leq kr$. (Note the irrelevance of the exponents $e_{i,\ell}$'s.) 
So, if $\Psi$ is faithful to the set
$\mathcal{P} := \{ P_{i, \ell}, \partial_j P_{i, \ell} : 1 \leq i, j \leq
r, 1 \leq \ell \leq kr \}$ and the singleton sets $\{P_{i, \ell}\}$ for
$1 \leq i \leq r$, $kr +1 \leq \ell \leq d$, then
$\Psi(J_{\mathbf{x}_r}(\mathbf{T}_r)) \neq 0$. Observe that the
polynomials in $\mathcal{P}$ and the singleton sets are (zeroth and
first order) derivatives of the gates at level $3$, and further these
sets involve (the derivatives of) \emph{disjoint} groups of level-$3$
gates. This disjointness feature ensures that the number of such sets
is at most $s$. Thus, we have reduced the problem of constructing a
faithful map $\Phi$ for $\mathbf{T}$ (gates at level $2$) to the
problem of constructing a map $\Psi$ that is faithful to at most $s$
many sets each containing derivatives of gates at the third
level. Now, the idea is to carry forward this argument recursively to
deeper levels: In the next level of the recursion we reduce the
problem to constructing a map that is faithful to at most $s$ sets
containing (zeroth, first and second order) derivatives of disjoint
groups of gates at level $4$, and so on. Eventually, the recursion
reaches the level of the sparse polynomials (the leaf nodes) where a
faithful map can be constructed using ideas from \cite{KS01}.

Let us formalize this proof idea. For any \emph{multiset} of variables
$S$, let $\Delta_S f$ denote the partial derivative of $f$ with
respect to the variables in $S$ (including repetitions, as $S$ is a
multiset). Let $\text{var}(S)$ denote the set of distinct variables in
$S$.

\vspace{-0.04in}
\begin{lemma}[Gcd trick]\label{lem:derivative-content}
  Let $G$ be any gate in $C$ and $S_1,\cdots, S_w$ be multisets of
  variables. Then there exists another occur-$k$ formula $G'$ for which, 
  the vector of polynomials $\inparen{\Delta_{S_1} G,
    \cdots, \Delta_{S_w} G} = V_G \cdot \inparen{\Delta_{S_1}G',
    \cdots, \Delta_{S_w}G'}$ such that
  \begin{enumerate}

    \vspace{-0.08in}
  \item If $G$ is a $+$ gate then $G'$ is also a $+$ gate whose
    children consist of at most $k \cdot
    \abs{\cup_{i=1}^{w}{\emph{var}(S_i)}}$ of the children of $G$, and
    $V_G = 1$.

    \vspace{-0.08in}
  \item If $G$ is a $\times\curlywedge$ gate, then $G'$ is also a $\times\curlywedge$ gate
    whose children consist of at most $k \cdot
    \abs{\cup_{i=1}^{w}{\emph{var}(S_i)}}$ of the children of $G$, and $V_G = G/G'$. 

  \end{enumerate}
  \vspace{-0.05in}
  Further, the gates constituting $G'$ and $V_G$ are disjoint. \\
  {\em Proof: Use properties of derivation and occur-$k$, see Appendix \ref{sec:app_hs-occ}.}
\end{lemma}

\vspace{-0.04in}
\noindent We say that a map is faithful to a collection of sets if it
is faithful to every set in the collection. Going by the `proof idea',
suppose at the $\ell$-th level of the recursion we want to construct a
$\Psi_{\ell}$ that is faithful to a collection of (at most) $s$ sets
of polynomials, each set containing at most $r_{\ell}$ partial
derivatives (of order up to $c_{\ell}$) of the gates at level
$\ell$. Moreover, the sets involve derivatives of disjoint groups of
gates. To begin with: $\ell = 2$ and we wish to construct a $\Psi_2$ 
that is faithful to just one set $\mathbf{T}$, so $r_2 \leq k$
and $c_2 = 0$. The next lemma captures the evolution of the recursion.

\vspace{-0.04in}
\begin{lemma}[Evolution via factoring]\label{lem:descent-jacobian}
  Let $\mathcal{U}$ be a set of $r_\ell$ derivatives (of orders up to $c_{\ell}$) of gates
  $\mathcal{G}_{\mathcal{U}}$ at level $\ell$, and $\mathcal{U}'$ be a
  transcendence basis of $\mathcal{U}$. Any $\abs{\mathcal{U}'} \times
  \abs{\mathcal{U}'}$ minor of
  $\mathcal{J}_{\mathbf{x}}(\mathcal{U}')$ is of the form
  $\prod_{i}{V_i^{e_i}}$, where $V_i$'s are polynomials in at most
  $r_{\ell+1} := (c_{\ell}+1) \cdot 2^{c_{\ell}+1} k \cdot
  {r_{\ell}}^2$ many derivatives (of order up to $c_{\ell+1} :=
  c_{\ell} + 1$) of disjoint groups of children of
  $\mathcal{G}_{\mathcal{U}}$. \\
  {\em Proof: Apply the gcd trick for each $G\in\mathcal{G}_{\mathcal{U}}$, 
  see Appendix \ref{sec:app_hs-occ}.}
\end{lemma}

\vspace{-0.04in}
\noindent Let $\mathcal{C}_\ell$ denote the collection of sets for
which we want to construct a faithful map $\Psi_\ell$ at the $\ell$-th
level of the recursion. The collection $\mathcal{C}_{\ell+1}$ is
formed from $\mathcal{C}_\ell$ using the above lemma: $V_i$ is a
polynomial in a set of derivatives of gates at the $(\ell+1)$-th level
- denote this set of derivatives by $\text{Elem}(V_i)$ - then
$\mathcal{C}_{\ell+1}$ consists of the sets $\text{Elem}(V_i)$ as
$\mathcal{U}$ varies over all the sets in the collection
$\mathcal{C}_{\ell}$. It follows from the lemma that the groups of
gates whose derivatives form the different $\text{Elem}(V_i)$'s are
disjoint and therefore $\abs{\mathcal{C}_{\ell+1}} \leq s$. Using
Lemma~\ref{lem:composition-lemma} \& \ref{lem:descent-jacobian}, we
can lift a map $\Psi_{\ell+1}$ to construct $\Psi_\ell$.

\vspace{-0.03in}
\begin{corollary}\label{cor:descent-faithful}
  If $\Psi_{\ell +1}$ is faithful to $\mathcal{C}_{\ell+1}$ then
  $\Psi_\ell: x_i \mapsto \inparen{\sum_{j=1}^{r_\ell} y_{j,\ell}
    \cdot (t_{\ell})^{ij}} + \Psi_{\ell + 1}(x_i)$ is faithful to
  $\mathcal{C}_\ell$, where
  $\inbrace{y_{1,\ell},\cdots,y_{r_{\ell},\ell},t_{\ell}}$ is a fresh
  set of variables.
\end{corollary}

\vspace{-0.08in}
\begin{proofof}{Theorem~\ref{thm:dDkrPIT}}
  Unfolding the recursion, we eventually reach the level of the sparse
  polynomials at depth $D-2$ and are required to construct a map
  $\Psi_{D-2}$ that is faithful to a collection $\mathcal{C}_{D-2}$ of
  at most $s$ sets of derivatives of sparse polynomials, each set
  containing at most $r_{D-2}$ elements. Using the relation between
  $r_{\ell+1}$ and $r_\ell$ from Lemma \ref{lem:descent-jacobian}, it
  is easy to bound $r_{D-2}$ by $R = (2k)^{2D\cdot 2^D}$. Let
  $\mathcal{U} \in \mathcal{C}_{D-2}$ with transcendence basis
  $\mathcal{U}'$. Any $\abs{\mathcal{U}'} \times \abs{\mathcal{U}'}$
  minor of $\mathcal{J}_{\mathbf{x}}(\mathcal{U}')$ is a sparse
  polynomial with sparsity bounded by $s^R$ and degree bounded by
  $sR$. Using \cite{KS01}, the nonzeroness of this determinant is
  maintained by one of the maps $\Phi_p: x_i \mapsto u^{(sR+1)^i \bmod
    p}$ as $p$ varies from $1$ to a fixed $\poly(s^R)$. Since
  $\abs{\mathcal{C}_{D-2}} \leq s$, one of the maps $\Phi_p$, $1 \leq
  p \leq s \cdot \poly(s^R)$, preserves the rank of the Jacobian of
  all $\mathcal{U}$ in $\mathcal{C}_{D-2}$ - fix such a
  $\Phi_p$. Finally, $\Psi_{D-2}: x_i \mapsto \sum_{j=1}^{R}{y_{j,D-2}}
   t_{D-2}^{i j} + \Phi_p(x_i)$ is faithful to
  $\mathcal{C}_{D-2}$. Now lift this map $\Psi_{D-2}$ to $\Psi_2$ that
  is faithful to $\mathbf{T}$ using Corollary
  \ref{cor:descent-faithful}. The map $\Psi_2$ reduces the number of
  variables to $O(R)$ and hence an application of  \cite{S80, Z79, DL78} lemma
   leads to a hitting-set generator with running time
  poly$(s^R)$. For the Jacobian criterion to work we need
  $\mathrm{char}(\F) = 0$ or $> s^R$.
\end{proofof}

\vspace{-0.2in}
\subsection{Restriction to the case of  depth-$4$} \label{sec:depth4occurkPIT}

\vspace{-0.1in}
\begin{proofof}{Theorem~\ref{thm:d4krPIT}}
  Let $C = \sum_{i=1}^{k}{T_i}$ be a depth-$4$ occur-$k$ formula,
  where $T_i = \prod_{j=1}^{d}{P_{ij}^{e_{ij}}}$, $P_{ij}$'s are
  sparse polynomials. The discussion at the beginning of this section
  justifies the assumption that top fanin is $k$. Once again,
  assuming $\mathbf{T}_r$ to be a transcendence basis of $\mathbf{T}$,
  we need to design a $\Psi$ such that
  $\Psi(J_{\mathbf{x}_r}(\mathbf{T}_r)) \neq 0$. Let us count the
  number of $P_{ij}$'s that depend on the variables $\mathbf{x}_r$,
  the remaining $P_{ij}^{e_{ij}}$'s can be taken out common from every
  row of $\mathcal{J}_{\mathbf{x}_r}(\mathbf{T}_r)$ while computing
  its determinant - this is the first `taking common' step.  Let
  $c_{i\ell}$ be the number of $P_{ij}$'s present in $T_i$ that depend
  on $x_\ell$. The total number of sparse polynomials depending on
  $\mathbf{x}_r$ is therefore $\sum_{1\le i,\ell\le r} c_{i\ell}$. From the
  condition of occur-$k$, $\sum_{i} c_{i\ell} \leq k$ and hence
  $\sum_{i,\ell} c_{i\ell} \leq rk \leq k^2$.  Let $c_i := \sum_j
  c_{ij}$, be the number of $\mathbf{x}_r$-dependent $P_{ij}$'s
  present in $T_i$. For an $\mathbf{x}_r$-dependent $P_{ij}$, we can
  also take $P_{ij}^{e_{ij}-1}$ common from the $i$-th row of
  $\mathcal{J}_{\mathbf{x}_r}(\mathbf{T}_r)$ - call this the second
  `taking common' step. The sparsity of every entry of the $i$-th row
  of the residual matrix $M$ - after the two `taking common' steps -
  is bounded by $c_i s^{c_i}$, where $s$ is the size of $C$. Thus,
  $\det(M)$ has sparsity at most $r!\cdot \prod_{i=1}^{r} {c_i
    s^{c_i}} = s^{O(k^2)}$, which implies that
  $J_{\mathbf{x}_r}(\mathbf{T}_r)$ is a product of at most $s+1$
  powers of sparse polynomials, each of whose sparsity is bounded by
  $s^{O(k^2)}$ and degree bounded by $sk$. As argued before, use
  \cite{KS01} along with Lemma \ref{lem:composition-lemma} to
  construct a hitting-set for $C$ in time $s^{O(k^2)}$ (assuming
  $\mathrm{char}(\F) = 0$ or $> s^{2k}$).
\end{proofof}

\vspace{-0.3in}
\section{Lower bounds for the immanant} \label{sec:lowerbounds}

\vspace{-0.1in} 
For convenience, we prove the lower bounds for 
$\text{Det}_n$ - determinant of an $n \times n$ matrix $M = (x_{ij})$ -
assuming zero characteristic. 
All our arguments apply to $\text{Imm}_{\chi}(M)$ for any character $\chi$
(Appendix \ref{sec:app_imm}). The following two lemmas are at the
heart of our approach to proving lower bounds. Let $\mathbf{x} :=
\{x_{ij}: 1 \leq i,j \leq n \}$ and $\mathbf{T} := \{ T_1, \ldots,
T_m\}$, where $T_i$'s are polynomials in $\F[\mathbf{x}]$. (See 
Appendix \ref{sec:app_lbnd} for the proofs.)

\vspace{-0.08in}
\begin{lemma}\label{lem:diag-minor-sum}
  Suppose $\emph{Det}_n = C(T_1, \ldots, T_m)$, where $C$ is any
  circuit and let $\mathbf{T}_r = \{ T_1, \ldots, T_r\}$ be a
  transcendence basis of $\mathbf{T}$ with $r < n$. Then, there exist
  a set of $r+1$ variables $\mathbf{x}_{r+1} \subset \mathbf{x}$ and
  an equation $\sum_{i=1}^{r+1} c_if_i \cdot M_i = 0$ such that
  $M_i$'s are distinct first order principal minors of $M$, $f_i$'s
  are distinct $r\times r$ minors of
  $\mathcal{J}_{\mathbf{x}_{r+1}}(\mathbf{T}_r)$, not all $f_i$'s are
  zero, and $c_i \in\F^*$. 
\end{lemma}

\vspace{-0.15in}
\begin{lemma}\label{lem:sparse-comb-minors}
  If $M_1,\cdots, M_t$ are distinct first order principal minors of
  $M$ and $\sum_{i=1}^{t} f_i \cdot M_i = 0$ (not all $f_i$'s are
  zero) then the total sparsity of the $f_i$'s is at least $2^{n/2 -
    t}$. 
\end{lemma}

\vspace{-0.15in}
\subsection{Lower bound on depth-$4$ occur-$k$ formulas}

\vspace{-0.1in}
\begin{proofof}{Theorem \ref{thm:lowerbnd-depth4-occurk}}
  Let $C$ be a depth-$4$ occur-$k$ formula of size $s$ that computes
  $\text{Det}_n$. Since $\text{Det}_n$ is irreducible we can assume a top
  $+$ gate in $C$. Then $\tilde{C} := C(x_{11}+1, x_{12}, \ldots,
  x_{nn}) - C(\mathbf{x})$ is a depth-$4$ occur-$2k$ formula of size
  at most $2s^2$ and top fanin bounded by $2k$ (similar argument as at the 
  beginning of Section
  \ref{sec:depthDoccurkPIT}). Moreover, $\tilde{C}$ computes the minor
  of $M$ with respect to $x_{11}$ which is essentially
  $\text{Det}_{n-1}$. By reusing symbols, assume that $C$ is a
  depth-$4$ occur-$k$ formula with top fanin bounded by $k$, and $C$
  computes $\text{Det}_n$.

  Let $C = \sum_{i=1}^{k}{T_i} = \text{Det}_n$, where $T_i =
  \prod_{j=1}^{d}{P_{ij}^{e_{ij}}}$, $P_{ij}$'s are sparse
  polynomials. Let $\mathbf{T}_r$ be a transcendence basis of
  $\mathbf{T} = \{T_1, \ldots, T_k\}$. By Lemma
  \ref{lem:diag-minor-sum}, we have an equation $\sum_{i=1}^{r+1} c_i
  f_i \cdot M_i = 0$ such that $f_i$'s are distinct $r \times r$
  minors of $\mathcal{J}_{\mathbf{x}_{r+1}}(\mathbf{T}_r)$ for some
  set of $r+1$ variables $\mathbf{x}_{r+1}$. Arguing in the same way
  as in the proof of Theorem~\ref{thm:d4krPIT} (in Section
  \ref{sec:depth4occurkPIT}), we can throw away certain common terms
  from the minors $f_i$'s and get another equation
  $\sum_{i=1}^{r+1}{g_i M_i} = 0$, where the sparsity of each $g_i$ is
  $s^{O(k^2)}$. If we apply Lemma \ref{lem:sparse-comb-minors} on this
  equation, we get our desired result.
\end{proofof}

\vspace{-0.2in}
\subsection{Lower bound on circuits generated by $\Sigma\Pi$ polynomials}

\vspace{-0.1in}
\begin{proofof}{Theorem \ref{thm:lowerbnd-sparse-algRank}}
  In Lemma~\ref{lem:diag-minor-sum}, take the $T_i$'s to be sparse
  polynomials with sparsity bounded by $s$. Then, in the equation
  $\sum_{i=1}^{r+1} c_if_i \cdot M_i = 0$, each $f_i$ has sparsity
  $s^{O(r)}$. Finally, apply Lemma~\ref{lem:sparse-comb-minors} to
  obtain the desired lower bound.
\end{proofof}

\vspace{-0.2in}
\subsection{Lower bound on circuits generated by $\Pi\Sigma$ polynomials}
\vspace{-0.1in}
\begin{proofof}{Theorem \ref{thm:lowerbound-depth3}}
  Let $\mathbf{T} = \inbrace{T_1,\cdots, T_m}$ be products of linear
  polynomials such that $C(T_1,\cdots,T_m) = \Det _n$ with
  $\mathbf{T}_k = \inbrace{T_1,\cdots, T_k}$ being a transcendence
  basis. By Lemma~\ref{lem:diag-minor-sum}, we get $\sum_{i=1}^{k+1} c_i f_i M_i =
  0$ where the $f_i$'s are $k\times k$ minors of
  $\mathcal{J}_{\mathbf{x}_{k+1}}(\mathbf{T}_k)$ and wlog $f_1\ne0$. 
  Like in Section~\ref{sec:PIT-d3}, we
  can rewrite this equation in the form $H_0 := T \cdot \sum_{L}
  \alpha_{L}(\mathbf{M}_{k+1})/\ell_1\cdots\ell_k = 0$ where
  $\alpha_L(\mathbf{M}_{k+1}) := \sum_{i=1}^{k+1} \alpha_{L,i} M_i$ is
  an $\F$-linear combination of $\mathbf{M}_{k+1} :=
  \inbrace{M_1,\cdots, M_{k+1}}$. Observe that $H_0$ is a sum of
  products of linear polynomials, with `coefficients' being $\F$-linear
  combinations of $\mathbf{M}_{k+1}$. And since $f_1\neq 0$, the
  `coefficient' of $M_1$ in $H_0$ is a nonzero depth-$3$ circuit.

  The idea is to apply a similar treatment as in
  Section~\ref{sec:PIT-d3} to evolve $H_0$. The invariant that shall
  be maintained is that the coefficient of $M_1$ (modulo some linear
  polynomials), which is a depth-$3$ circuit, would stay
  nonzero. This would finally yield a non-trivial linear combination
  $\alpha_L(\mathbf{M}_{k+1}) = 0\bmod{\boldsymbol\ell_k}$ whence we
  can apply the following lemma. (See Appendix~\ref{sec:app_lbnd}.)

  \vspace{-0.04in}	
  \begin{lemma}\label{lem:det-ind-modulo}
    If $M_1,\cdots, M_t$ are distinct first order principal minors of
    $M$ and $\sum_{i=1}^{t}\alpha_i M_i = 0\bmod \boldsymbol\ell_k$ (not all
    $\alpha_i = 0$) for independent linear polynomials
    $\boldsymbol\ell_k$, then $t + k \ge n$. 
  \end{lemma}
  
  \vspace{-0.04in}
  \noindent
  Formally, define the \emph{content} of a circuit $H = T\sum_L
  \alpha_L(\mathbf{M}_{k+1})/\ell_1\cdots\ell_k$ as $\text{cont}(H) :=
  \gcd_L\inbrace{\frac{T}{\ell_1\cdots\ell_k}}$, and define
  $\text{sim}(H) := H/\text{cont}(H)$. Let $\text{sim}(H_0)$ have the
  form $F_0\sum_L \alpha_L(\mathbf{M}_{k+1})/\ell_1\cdots\ell_k$. The
  coefficient of $M_1$ in the above expression is a nonzero depth-$3$
  circuit, whose degree is $|\mathcal{L}(H_0)| - k$. Therefore 
  by Chinese remaindering, $\exists\ell_1\in \mathcal{L}(H_0)$ such 
  that this coefficient is
  nonzero modulo $\ell_1$. Hence, we can define $H_1 :=
  \text{sim}(H_0)\bmod\ell_1$ which has the form $H_1 = F_0/\ell_1\cdot
  \sum_{L\ni \ell_1} \alpha_L(\mathbf{M}_{k+1})/\ell_2\cdots\ell_k = 0
  \bmod{\ell_1}$. And like in Section~\ref{sec:PIT-d3}, the above
  equation can be rewritten by replacing a variable occuring in
  $\ell_1$ by a suitable linear combination of the rest. Thus, we may
  write $H_1 = F_1\sum_L\alpha_L(\mathbf{M}_{k+1}\bmod{\ell_1})/
  \ell_2\cdots\ell_k = 0$, and maintaining
  the invariant that the coefficient of $M_1\bmod\ell_1$ is nonzero. Repeating
  this argument, we eventually obtain $H_k := F_k
  \cdot\alpha_L(\mathbf{M}_{k+1}\bmod{{\boldsymbol\ell}_k}) = 0$ while
  the coefficient of $M_1\bmod{\boldsymbol\ell}_k$ is nonzero. This
  implies that $\alpha_L(\mathbf{M}_{k+1}) = 0\bmod\boldsymbol\ell_k$
  is a non-trivial equation. And Lemma~\ref{lem:det-ind-modulo}
  asserts that this is not possible unless $2k+1\ge n$ or $k \ge (n-1)/2$.
\end{proofof}

\vspace{-0.25in}
\section{Conclusion}\label{sec-conclusion}
\vspace{-0.1in} 
We would like to note that the proof technique used to
show the lower bound for depth-$4$ occur-$k$ formulas can be
extended to prove an exponential lower bound for constant-depth constant-occur
 formulas if the following conjecture is true (see
Appendix~\ref{sec:app_cond_lbnd} for details).
 
\vspace{0.05in}
\noindent
{\bf `Determinant of immanants' conjecture: }
  Let $M = (x_{ij})$ be an $n\times n$ matrix, and let $x_i$ denote
  the $i$-th diagonal variable $x_{ii}$. Let $M'$ be a projection of
  $M$ by setting $c=o(n)$ of the variables in $M$ to constants. Suppose the
  elements $\mathbf{x}_k:=\{x_1,\cdots, x_k\}$, where $k$ is a constant independent of
  $n$, are partitioned into non-empty sets $\mathbf{S}_t:=\{S_1,\cdots, S_t\}$. Consider
  $\mathcal{M}(\mathbf{S}_t)$, the set of all $t^{\text{th}}$ order
  principal minors of $M'$, each by choosing a $t$-tuple $B\in
  S_1\times\cdots \times S_t$ as pivots. Over all possible choices of
  $B$, we get $m := |S_1|\cdots |S_t|$ many minors. Then for any set of
  diagonal variables $\mathbf{y}_m$ disjoint from $\mathbf{x}_k$, 
  $J_{\mathbf{y}_m}\inparen{\mathcal{M}(\mathbf{S}_t)}\ne0$.
%
%

\vspace{0.05in}
\noindent In (Jacobian) spirit, the conjecture states that the
$t^{\text{th}}$-order principal minors $N_{B_1}, \ldots, N_{B_m}$ are
algebraically independent, when $n$ is sufficiently large (say, $n>c+k
+ m$).

Spurred by the success of Jacobian in solving the
hitting-set problem for \emph{constant-trdeg} depth-$3$ circuits
and \emph{constant-occur} constant-depth formulas, one is naturally
inspired to investigate the strength of this approach against other
`constant parameter' models - the foremost of which is {\em constant top
fanin} depth-$4$ circuits (PIT even for fanin $2$?).

Another problem, which is closely related to hitting-sets
and lower bounds, is reconstruction of arithmetic circuits \cite[Chapter 5]{SY10}. 
There is a quasi-polynomial time
reconstruction algorithm \cite{KS09}, for a polynomial computed by a
depth-$3$ constant top fanin circuit, that outputs a depth-$3$
circuit with quasi-polynomial top fanin. 
Could Jacobian be used
as an effective tool to solve reconstruction problems? 
If yes, then it would further reinforce the versatility of this tool.


\newpage

\footnotetext[1]{Estimates the bit complexity of the hitting-set generator; 
constant factors not stressed (also in higher exponents).}
\footnotetext[2]{We assume a zero or large characteristic.}
\begin{figure}
\caption{Comparison with the earlier efficient hitting-sets}
\centering
{\scriptsize
\begin{tabular}{|c|c|c|c|c|}
  \hline
  \multicolumn{2}{|c|}{{\bf Previous best}} & \multicolumn{3}{|c|}{{\bf This paper}}\\\hline
  \emph{Model} & \emph{Running time\footnotemark[1]} & \emph{Extended Model\footnotemark[2]} & 
  \emph{Running time\footnotemark[1]} & {\em $\text{Imm}_n$ lower bound}\\\hline
  $\Sigma\Pi\Sigma(k)$ circuits: & $s^k$ & $C(T_1,\cdots, T_m) \stackrel{?}{=} 0$& \multirow{2}{*}{$s^k$} & \multirow{2}{*}{$\mathrm{trdeg}\inbrace{T_i}=\Omega(n)$}\\
  $T_1 + \cdots + T_k \stackrel{?}{=}0$ &\cite{SS11}  &  {\tiny poly-degree $C$ \& $\mathrm{trdeg}\inbrace{T_i}\leq k$} & & \\\hline
  $\Sigma\Pi\Sigma\Pi(k)$ & $s^{k^3}$& $\Sigma\Pi\Sigma\Pi$ & \multirow{2}{*}{$s^{k^2}$}& \multirow{2}{*}{$s = 2^{\Omega(n/k^2)}$} \\
  {\tt multilinear} circuits & \cite{SV11} & occur-$k$ formulas & &\\\hline
  depth-$D$, read-$k$ & $s^R$ & depth-$D$, occur-$k$& $s^R$ & $s = 2^{\Omega(n)}$\\
  {\tt multilinear} formulas & where $R = k^{k^2} + kD$& formulas & where $R = k^{2^D}$ & {\tiny for constant $k,D$} \\
  & \cite{AMV11}& & & {\tiny assuming Conjecture~\ref{conj:minor-conjecture}}\\\hline
  $C(f_1,\cdots, f_m) \stackrel{?}{=} 0$ & $s^k$ &  &  & \\
  {\tiny poly-degree $C$, $\Sigma\Pi$ circuits $f_i$'s}  & \cite{BMS11} & -- & -- & $s = 2^{\Omega(n/k)}$ \\
  {\tiny \& $\mathrm{trdeg}\inbrace{f_i}\leq k$} &  & & & \\\hline
\end{tabular}
}
\label{fig-1}
\end{figure}

\bibliography{references}


\appendix

\section{Missing Proofs}

\subsection{Preliminaries: Jacobian and faithful homomorphisms}\label{sec:app_prel}

\begin{reptheorem}{thm:faithful-pit}
  Let $\mathbf{f} = \inbrace{f_1,\cdots, f_m} \subset \F[\mathbf{x}]$ 
  and $\Phi$ be a homomorphism faithful to $\mathbf{f}$. For any polynomial
  $C\in\F[y_1,\cdots, y_m]$, $C(\mathbf{f}) = 0 \Leftrightarrow C(\Phi(\mathbf{f})) = 0$.
\end{reptheorem}
\begin{proof}
  Since $\Phi$ is faithful to $\mathbf{f}$, there is a transcendence
  basis (say, $f_1, \ldots, f_s$) of $\mathbf{f}$ such that
  $\Phi(f_1), \ldots, \Phi(f_s)$ is a transcendence basis of
  $\Phi(\mathbf{f})$. The function field $\mathbb{K} = \F(\mathbf{f})$
  essentially consists of elements that are polynomials in $f_{s+1},
  \ldots, f_m$ with coefficients from $\F(f_1, \ldots, f_s)$. Treating
  $C(\mathbf{f})$ as a nonzero element of $\mathbb{K}$, there is an
  inverse $Q \in \mathbb{K}$ such that $Q \cdot C = 1$. Since $Q$ is a
  polynomial in $f_{s+1}, \ldots, f_m$ with coefficients from $\F(f_1,
  \ldots, f_s)$, by clearing off the denominators of these
  coefficients in $Q$, we get an equation $\tilde{Q} \cdot C = P(f_1,
  \ldots, f_s)$, where $\tilde{Q}$ is a nonzero polynomial in
  $\mathbf{f}$ and $P$ is a nonzero polynomial in $f_1, \ldots,
  f_s$. Applying $\Phi$ to both sides of the equation, we conclude
  that $C(\Phi(\mathbf{f})) = \Phi(C(\mathbf{f})) \neq 0$, otherwise
  $P(\Phi(f_1), \ldots, \Phi(f_s)) = \Phi(P(f_1, \ldots, f_s)) = 0$
  which is not possible as $\Phi(f_1), \ldots, \Phi(f_s)$ are
  algebraically independent and $P$ is a nontrivial polynomial.
\end{proof}

\begin{lemma}[Vandermonde map]\label{lem:gabizon-raz}
Let $A$ be a $r\times n$ matrix with entries in a field $\F$, and 
let $t$ be an indeterminate. Then, $\mathrm{rank}_{\F(t)} 
\inparen{ A\cdot(t^{ij})_{i\in[n],j\in[r]} }$ $=$ $\mathrm{rank}_{\F} A$.
\end{lemma}
\begin{proof}
  Follows from Lemma 6.1 of \cite{GR05}. 
\end{proof}

\begin{corollary}\label{cor:gabizon-raz}
Let $V_1,\cdots, V_t$ be $k$-dimensional subspaces of linear 
polynomials in $\F[x_1,\cdots, x_n]$. For a constant $\alpha\in\F$, 
define the following linear homomorphism $\Psi_\alpha$ as
$$
\Psi_\alpha:x_i \mapsto \sum_{j=1}^k y_j \alpha^{ij}
$$
If $|\F| > tnk^2$, then there exists an $\alpha$ such that 
$\Psi_\alpha$ is an isomorphism on each of $V_1,\cdots, V_t$. 
\end{corollary}

\begin{replemma}{lem:composition-lemma}
  Let $\mathbf{f} \subset \F[\mathbf{x}]$ be a finite set of
  polynomials of degree at most $d$, 
  $\mathrm{trdeg}_{\F} \hspace{0.025in} \mathbf{f} \leq r$, and
  $\emph{char}{(\F)} = 0$ or $> d^r$. Let 
  $\Psi:\F[\mathbf{x}]\rightarrow \F[\mathbf{z}]$ be a homomorphism such that
  $\mathrm{rank}_{\F(\mathbf{x})} \mathcal{J}_{\mathbf{x}}(\mathbf{f})$ $=$
  $\mathrm{rank}_{\F(\mathbf{z})} \Psi(\mathcal{J}_{\mathbf{x}}(\mathbf{f}))$.
  
  Then, the map $\Phi:\F[\mathbf{x}]\rightarrow$ 
  $\F[\mathbf{z},t, y_1, \ldots, y_r]$ that maps, 
  for all $i$, $x_i \mapsto \inparen{\sum_{j=1}^r y_j t^{ij}} + \Psi(x_i)$ is 
  a homomorphism faithful to $\mathbf{f}$.
\end{replemma}
\begin{proof}
Wlog let $\mathrm{trdeg}_\F \mathbf{f}=r$, which then (by Jacobian criterion) 
is the rank of $\mathcal{J}_{\mathbf{x}}(\mathbf{f})$. We intend to show 
that the matrix $\mathcal{J}_{\mathbf{y}}(\Phi(\mathbf{f}))$ is of rank $r$,
which would imply (by Jacobian criterion) that $\mathrm{trdeg}_{\F(t,\mathbf{z})}$ 
$\Phi(\mathbf{f}) = r$.

Consider the projection $\mathcal{J}'$ of $\mathcal{J}_{\mathbf{y}}(\Phi(\mathbf{f}))$
obtained by setting $y_1=\cdots=y_r=0$. 
\begin{eqnarray*}
\mathcal{J}'= \insquare{\mathcal{J}_{\mathbf{y}}(\Phi(\mathbf{f}))}_{\mathbf{y}=\mathbf{0}} 
&=&
\insquare{ \Phi\inparen{ \mathcal{J}_{\mathbf{x}}(\mathbf{f}) }\cdot 
	\mathcal{J}_{\mathbf{y}}(\Phi(\mathbf{x})) }_{\mathbf{y}=\mathbf{0}}\hspace{0.2in}\text{(By chain rule)}\\
&=& 
\Psi\inparen{ \mathcal{J}_{\mathbf{x}}(\mathbf{f}) }\cdot 
	\mathcal{J}_{\mathbf{y}}(\Phi(\mathbf{x}))
\end{eqnarray*}
\noindent
Observe that the matrix $\mathcal{J}_{\mathbf{y}}(\Phi(\mathbf{x}))$ is exactly the 
Vandermonde matrix that is present in Lemma~\ref{lem:gabizon-raz}. Also,
$\Psi(\mathcal{J}_{\mathbf{x}}(\mathbf{f}))$ has entries in $\F(\mathbf{z})$, and 
by assumption has the same rank as $\mathcal{J}_{\mathbf{x}}(\mathbf{f})$. 
Hence, by Lemma~\ref{lem:gabizon-raz},
$$ \mathrm{rank}_{\F(t,\mathbf{z})} \mathcal{J}' =
\mathrm{rank}_{\F(t,\mathbf{z})} \inparen{\Psi(\mathcal{J}_{\mathbf{x}}(\mathbf{f}))\cdot
  \mathcal{J}_{\mathbf{y}}(\Phi(\mathbf{x})) }= 
  \mathrm{rank}_{\F(\mathbf{z})} \Psi(\mathcal{J}_{\mathbf{x}}(\mathbf{f})) = r.
$$
And since $\mathcal{J}'$ is just a projection of
$\mathcal{J}_{\mathbf{y}}(\Phi(\mathbf{f}))$, the rank of the latter must also be $r$. Hence,
$\Phi$ is indeed faithful. 
\end{proof}

\subsection{Hitting-set for constant transcendence degree depth-$3$ circuits}\label{sec:app_hs-d3}

\begin{replemma}{lem:path-cert-rational-sum}
 There exists independent linear polynomials $\inbrace{\ell_1,\cdots,
  \ell_k} \subseteq \mathcal{L}(H_0)$ such that $H_i \neq 0 \bmod
 (\ell_1, \ldots, \ell_i)$, $\forall i\in[k]$, and $H_k$ is a nonzero product of linear
 polynomials in $\mathcal{L}(H_0)$ modulo $(\ell_1,\cdots, \ell_k)$.
\end{replemma}
\begin{proof}
  The proof is by induction on $k$ and follows the sketch given while 
  defining sim$(\cdot)$. The degree of the nonzero polynomial 
  $\text{sim}(H_0)$
  is $\abs{\mathcal{L}(H_0)} - k$. By Chinese remaindering, there
  exists an $\ell_1 \in \mathcal{L}(H_0)$ such that $H_1 :=
  \text{sim}(H_0) \bmod \ell_1 \neq 0$. In the base case ($k=1$), it
  is easy to see that $H_1$ is a nonzero product of linear
  polynomials modulo $\ell_1$.  For any larger $k$, the depth-$3$
  polynomial $H_1$ has exactly the same form as $H_0$ but with $k-1$
  independent linear polynomials in the denominators. Induct on this
  smaller value $k-1$, keeping in mind that $\mathcal{L}(H_i)\subset
  \mathcal{L}(H_0)$ modulo $(\ell_1, \ldots, \ell_i)$.  
\end{proof}

Let $\mathcal{I}_i$ and
$\Psi(\mathcal{I}_i)$ denote the ideals generated by $\{\ell_1,
\ldots, \ell_i\}$ and $\{\Psi(\ell_1), \ldots, \Psi(\ell_i)\}$,
respectively. A rank-$(k+1)$ preserving $\Psi$ satisfies: $\ell
\neq 0 \bmod \mathcal{I}_i$ iff $\Psi(\ell) \neq 0 \bmod
\Psi(\mathcal{I}_i)$, for all  $1 \leq i \leq k$ and $\ell \in
\cup_{i=1}^{k}{L(T_i)}$.

\begin{reptheorem}{thm:varred}
  If $\Psi:\F[\mathbf{x}] \rightarrow \F[z_1,...,z_{k+1}]$ is a 
  rank-$(k+1)$ preserving map for $H_0$, then $\Psi(H_0) \neq 0$.
\end{reptheorem}
\begin{proof}
  Let $\{ \ell_1, \ldots, \ell_k\}$ be the certifying path of $H_0$ fixed
  above. The proof is by reverse induction on $k$: Assuming $\Psi(H_i)
  \neq 0 \bmod \Psi(\mathcal{I}_i)$, we show that $\Psi(H_{i-1}) \neq
  0 \bmod \Psi(\mathcal{I}_{i-1})$ for $k \geq i \geq 2$. The base
  case: By Lemma \ref{lem:path-cert-rational-sum}, $H_k$ is a
  \emph{nonzero} product of linear polynomials in $\mathcal{L}(H_0)$
  modulo $\mathcal{I}_k$, so by the definition of a rank-$(k+1)$
  preserving map, $\Psi(H_k) \neq 0 \mod \Psi(\mathcal{I}_k)$ (ideal
  generated by independent linear polynomials is an integral
  domain). By construction, $H_{i-1} = \text{cont}(H_{i-1}) \cdot
  \text{sim}(H_{i-1}) = \text{cont}(H_{i-1}) \cdot [q_i \ell_i + H_i]
  \mod \mathcal{I}_{i-1}$, for some polynomial $q_i$. Which means,
  $\Psi(H_{i-1}) = \Psi(\text{cont}(H_{i-1})) \cdot [\Psi(q_i)
  \Psi(\ell_i) + \Psi(H_i)] \bmod \Psi(\mathcal{I}_{i-1})$. If
  $[\Psi(q_i) \Psi(\ell_i) + \Psi(H_i)] = 0 \bmod
  \Psi(\mathcal{I}_{i-1})$, then $\Psi(H_i) = 0 \bmod
  \Psi(\mathcal{I}_i)$ which contradicts the induction
  hypothesis. Also, by Lemma \ref{lem:path-cert-rational-sum},
  $H_{i-1} \neq 0 \bmod \mathcal{I}_{i-1}$ implying that
  $\text{cont}(H_{i-1}) \neq 0 \bmod \mathcal{I}_{i-1}$. Since, $i\ge2$, the
  linear polynomials in the term $\text{cont}(H_{i-1})$ belong to
  $\mathcal{L}(H_0)$ modulo the ideal $\mathcal{I}_{i-1}$, once again
  by using the rank-$(k+1)$ preserving property of $\Psi$, we infer
  that $\Psi(\text{cont}(H_{i-1})) \neq 0 \bmod
  \Psi(\mathcal{I}_{i-1})$. Therefore, $\Psi(H_{i-1}) \neq 0 \bmod
  \Psi(\mathcal{I}_{i-1})$. Finally, to obtain $\Psi(H_0) \neq 0$ from
  $\Psi(H_1) \neq 0 \bmod \Psi(\mathcal{I}_1)$, use the same argument as
  above and that $\Psi(\ell) \neq 0$ for every $\ell \in
  \cup_{i=1}^{k}{L(T_i)}$.
\end{proof}

\subsection{Hitting-set for constant-depth constant-occur formulas}\label{sec:app_hs-occ}

\begin{replemma}{lem:derivative-content}
Let $G$ be any gate in $C$ and $S_1,\cdots, S_w$ be multisets of
  variables. Then there exists another occur-$k$ formula $G'$ for which, 
  the vector of polynomials $\inparen{\Delta_{S_1} G,
    \cdots, \Delta_{S_w} G} = V_G \cdot \inparen{\Delta_{S_1}G',
    \cdots, \Delta_{S_w}G'}$ such that
  \begin{enumerate}

    \vspace{-0.08in}
  \item If $G$ is a $+$ gate then $G'$ is also a $+$ gate whose
    children consist of at most $k \cdot
    \abs{\cup_{i=1}^{w}{\emph{var}(S_i)}}$ of the children of $G$, and
    $V_G = 1$.

    \vspace{-0.08in}
      \item If $G$ is a $\times\curlywedge$ gate, then $G'$ is also a $\times\curlywedge$ gate
    whose children consist of at most $k \cdot
    \abs{\cup_{i=1}^{w}{\emph{var}(S_i)}}$ of the children of $G$, and $V_G = G/G'$. 

  \end{enumerate}
  \vspace{-0.05in}
  Further, the gates constituting $G'$ and $V_G$ are disjoint. 
\end{replemma}

\begin{proof} 
  \begin{enumerate}
  \item Suppose $G = H_1 + \cdots + H_m$. Then at most $k\cdot
    |\union \text{var}(S_i)|$ of its children depend on the variables present in
    $\union \text{var}(S_i)$; let $G'$ be the sum of these children. Then,
    $\Delta_{S_i}G = \Delta_{S_i}G'$ as the other gates are
    independent of the variables in $\union S_i$.
   
  \item Suppose $G = H_1^{e_1} \cdots H_m^{e_m}$. Since $G$ is a gate
    in an occur-$k$ formula, at most $k\cdot |\union \text{var}(S_i)|$ of the
    $H_i$'s depend on the variables in $\union S_i$; call these
    $H_1,\cdots, H_t$. Let $G' := H_1^{e_1}\cdots H_t^{e_t}$ and $V_G :=
    G/G'$. Then, $\Delta_{S_i}G = V_G\cdot \Delta_{S_i}G'$ as claimed.\qedhere 
  \end{enumerate}
\end{proof}

\begin{replemma}{lem:descent-jacobian}
  Let $\mathcal{U}$ be a set of $r_\ell$ derivatives (of orders up to $c_{\ell}$) of gates
  $\mathcal{G}_{\mathcal{U}}$ at level $\ell$, and $\mathcal{U}'$ be a
  transcendence basis of $\mathcal{U}$. Any $\abs{\mathcal{U}'} \times
  \abs{\mathcal{U}'}$ minor of
  $\mathcal{J}_{\mathbf{x}}(\mathcal{U}')$ is of the form
  $\prod_{i}{V_i^{e_i}}$, where $V_i$'s are polynomials in at most
  $r_{\ell+1} := (c_{\ell}+1) \cdot 2^{c_{\ell}+1} k \cdot
  {r_{\ell}}^2$ many derivatives (of order up to $c_{\ell+1} :=
  c_{\ell} + 1$) of disjoint groups of children of
  $\mathcal{G}_{\mathcal{U}}$.
\end{replemma}
\begin{proof}
  Let $G \in \mathcal{G}_{\mathcal{U}}$ be a gate at level $\ell$ and
  $\{U_1, \ldots, U_{e_G}\} \subset \mathcal{U}'$ be the set of all
  the derivatives of $G$ present in $\mathcal{U}'$. Fix any
  $\abs{\mathcal{U}'} \times \abs{\mathcal{U}'}$ sub-matrix $M$ of
  $\mathcal{J}_{\mathbf{x}}(\mathcal{U}')$. Consider the $e_G$ rows of
  $M$ that contain the derivatives of $U_1, \ldots, U_{e_G}$. These
  rows together contain a total of $w := e_G \cdot \abs{\mathcal{U}'}$
  elements that are up to $(c_{\ell} + 1)$-order derivatives of $G$; view 
  all the elements of these $e_G$ rows as a single vector 
  $\inparen{\Delta_{S_1} G, \cdots, \Delta_{S_w} G}$ and apply 
  Lemma \ref{lem:derivative-content} to
  express it as $V_G \cdot \inparen{\Delta_{S_1}G', \cdots,
    \Delta_{S_w}G'}$. Verify that
  $\abs{\cup_{i=1}^{w}{\text{var}(S_i)}} \leq e_G \cdot c_{\ell} +
  \abs{\mathcal{U}'} \leq e_G \cdot c_{\ell} + r_\ell$. So, in
  $\det(M)$ we can take $V_G$ common from each of these $e_G$ rows
  such that the elements present inside the determinant are of the
  form $\Delta_{S_i}G'$, where $G'$ has at most $k(e_G c_\ell +
  r_\ell)$ children.

  Since $\abs{S_i} \le c_\ell + 1$, at most $k(c_\ell + 1)$ children of
  $G'$ depend on $\text{var}(S_i)$. If $G'$ is a $+$ gate, then
  $\Delta_{S_i}G'$ is the sum of the derivatives of at most $k(c_\ell +
  1)$ of its children (that depend on $\text{var}(S_i)$). If $G'$ is a
  $\times\curlywedge$ gate computing $H_1^{e_1}\cdots H_t^{e_t}$ (where $t\leq
  k(e_G c_\ell + r_\ell)$), then $\Delta_{S_i}G'$ is a polynomial
  combination of the $H_i$'s and
  $\inbrace{\Delta_{T}H_j}_{\emptyset\neq T\subseteq S_i}$ for each
  $H_j$ depending on $\text{var}(S_i)$.  Hence in either case,
  $\Delta_{S_i}G'$ is a polynomial in the children of $G'$ and their
  at most $(2^{c_\ell+1} - 1)\cdot k(c_\ell+1)$ many derivatives (of order between 
  one and $(c_\ell+1)$).

  Summing over all the $w$ elements
  $\Delta_{S_i}G'$, the elements of the $e_G$ rows of $M$ are
  polynomials in at most $k(e_G c_\ell + r_\ell) + w \cdot
  (2^{c_\ell+1} - 1)k(c_\ell+1)=$ $k(e_G c_\ell + r_\ell) + e_G \cdot \abs{\mathcal{U}'} \cdot
  (2^{c_\ell+1} - 1)k(c_\ell+1)$  
  derivatives of the children of $G'$. Going over all $G
  \in \mathcal{G}_{\mathcal{U}}$, $\det(M)$ can be expressed as a
  product $\prod_{G \in \mathcal{G}_{\mathcal{U}}}{V_G^{e_G}}$ and a
  polynomial $V$ in at most $k(r_\ell c_\ell + r_\ell^2) + r_\ell^2\cdot(2^{c_\ell+1} - 1)k(c_\ell+1)\le$
  $(c_{\ell}+1)2^{c_{\ell}+1} k {r_{\ell}}^2$
  derivatives (of order up to $c_\ell + 1$) of a group of gates in level
  $\ell + 1$. Further, the groups of gates whose derivatives
  constitute the $V_G$'s and $V$ are mutually disjoint (by Lemma
  \ref{lem:derivative-content}).
\end{proof}

\subsection{Lower bounds for the immanant}\label{sec:app_lbnd}

\begin{replemma}{lem:diag-minor-sum}
  Suppose $\emph{Det}_n = C(T_1, \ldots, T_m)$, where $C$ is any
  circuit and let $\mathbf{T}_r = \{ T_1, \ldots, T_r\}$ be a
  transcendence basis of $\mathbf{T}$ with $r < n$. Then, there exist
  a set of $r+1$ variables $\mathbf{x}_{r+1} \subset \mathbf{x}$ and
  an equation $\sum_{i=1}^{r+1} c_if_i \cdot M_i = 0$ such that
  $M_i$'s are distinct first order principal minors of $M$, $f_i$'s
  are distinct $r\times r$ minors of
  $\mathcal{J}_{\mathbf{x}_{r+1}}(\mathbf{T}_r)$, not all $f_i$'s are
  zero, and $c_i \in\F^*$. 
\end{replemma}
\begin{proof}
    In a column of a Jacobian matrix $\mathcal{J}_{\mathbf{x}}(\cdot)$,
  all the entries are differentiated with respect to a variable $x$,
  we will say that the column is \emph{indexed} by $x$. Let
  $\mathbf{T}_r = \inbrace{T_1,\cdots, T_r}$ be a transcendence basis
  of $\mathbf{T}$. Amongst the nonzero $r \times r$ minors of
  $\mathcal{J}_{\mathbf{x}}(\mathbf{T}_r)$ (they exist by Jacobian criterion), 
  pick one (call the matrix
  associated with the minor, $N$) that maximizes the number of
  diagonal variables $\{x_{ii}: 1 \leq i \leq n\}$ indexing the
  columns of $N$. Let $S$ denote the set of variables indexing the
  columns of $N$. Since $r < n$, there exists a diagonal variable
  $x_{jj} \notin S$. Consider the $(r+1)\times(r+1)$ minor of
  $\mathcal{J}_{\mathbf{x}}(\{\text{Det}_n\} \cup \mathbf{T}_r)$
  corresponding to the columns indexed by $S' := S \union
  \inbrace{x_{jj}}$ - call the associated $(r+1) \times (r+1)$ matrix
  $\tilde{N}$. Since, $\text{Det}_n = C(\mathbf{T})$, the polynomials
  $\text{Det}_n$ and $T_1, \ldots, T_r$ are algebraically dependent
  and hence $\det(\tilde{N})=0$. Expanding $\det(\tilde{N})$ along the
  first row of $\tilde{N}$, which contains signed first order minors
  (cofactors) of $M$, we have an equation $\sum_{i=1}^{r+1}{c_i f_i
    M_i =0}$, where $M_i$'s are distinct minors of $M$, $f_i$'s are
  distinct $r \times r$ minors of $\mathcal{J}_{S'}(\mathbf{T}_r)$,
  and $c_i \in\F^*$. If $M_i$ is the principal minor of $M$
  with respect to the variable $x_{jj}$ then $f_i = \det(N) \neq 0$
  (by construction). 
  
  It suffices to show that if $M_i$ is a
  non-principal minor of $M$ then $f_i = 0$. Consider any
  non-principal minor $M_i$ in the above sum, say it is the minor of
  $M$ with respect to $x_{k \ell}$. The corresponding $f_i$ is
  precisely the $r\times r$ minor of $\mathcal{J}_{S'}(\mathbf{T}_r)$
  with respect to the columns $S'\setminus\inbrace{x_{k \ell}}=$ 
  $\inparen{S\setminus\inbrace{x_{k \ell}}} \union\inbrace{x_{jj}}$. Hence, by 
  the maximality assumption on the number
  of diagonal elements of $M$ in $S$, $f_i = 0$.
\end{proof}

\begin{replemma}{lem:sparse-comb-minors}
  If $M_1,\cdots, M_t$ are distinct first order principal minors of
  $M$ and $\sum_{i=1}^{t} f_i \cdot M_i = 0$ (not all $f_i$'s are
  zero) then the total sparsity of the $f_i$'s is at least $2^{n/2 -
    t}$.
\end{replemma}
\begin{proof}
  The proof is by contradiction. The idea is to start with the
  equation $\sum_{i=1}^{t}{f_i M_i} = 0$ and apply two steps -
  \emph{sparsity reduction} and \emph{fanin reduction} -
  alternatively, till we arrive at a contradiction in the form of an
  equation $f_j \cdot M_j = 0$, where neither $f_j$ nor $M_j$ is zero
  if the total sparsity of the $f_i$'s is less than $2^{n/2 -
    t}$. With an equation of the form $\sum_{i=1}^{\tau}{g_i N_i} =
  0$, we associate four parameters $\tau,$
  {\footnotesize$\mathcal{S}$}, $\eta$ and $c$. These parameters are
  as follows: $\tau$ is called the \emph{fanin} of the equation,
  {\footnotesize$\mathcal{S}$} is the total sparsity of the $g_i$'s
  (we always assume that not all the $g_i$'s are zero), every $N_i$ is
  a distinct first order principal minor of a symbolic $\eta \times
  \eta$ matrix $N = (x_{ij})$, and $c$ is the maximum number of
  entries of $N$ that are set as constants. To begin with, $g_i = f_i$
  and $N_i = M_i$ for all $1 \leq i \leq t$, so $\tau = t$,
  {\footnotesize$\mathcal{S}$} = $s$ (the total sparsity of the
  $f_i$'s), $\eta = n$, $N=M$ and $c=0$. In the `sparsity reduction'
  step, we start with an equation $\sum_{i=1}^{\tau}{g_i N_i} = 0$,
  with parameters $\tau,$ {\footnotesize$\mathcal{S}$}, $\eta$, $c$
  and arrive at an equation $\sum_{i=1}^{\tau'}{g_i' N_i'} = 0$ with
  parameters $\tau',$ {\footnotesize$\mathcal{S'}$}, $\eta'$, $c'$
  such that $\tau' \leq \tau$, {\footnotesize$\mathcal{S'}$} $\leq$
  {\footnotesize$\mathcal{S}$}/2, $\eta -1 \leq \eta' \leq \eta$, and
  $c' \leq c+1$. In the `fanin reduction' step, we start with an
  equation $\sum_{i=1}^{\tau}{g_i N_i} = 0$, with parameters $\tau,$
  {\footnotesize$\mathcal{S}$}, $\eta$, $c$ and arrive at an equation
  $\sum_{i=1}^{\tau'}{g_i' N_i'} = 0$ with parameters $\tau',$
  {\footnotesize$\mathcal{S'}$}, $\eta'$, $c'$ such that one of the
  two cases happens - Case $1$: $\tau' \leq \tau - 1$,
  {\footnotesize$\mathcal{S'}$} $\leq$ {\footnotesize$\mathcal{S}$},
  $\eta' = \eta - 1$, and $c' = c$; Case $2$: $\tau' = 1$,
  {\footnotesize$\mathcal{S'}$} $\leq$ {\footnotesize$\mathcal{S}$},
  $\eta' = \eta$, and $c' \leq c+\tau$.

  Naturally, starting with $\sum_{i=1}^{t}{f_i M_i} = 0$, the
  `sparsity reduction' step can only be performed at most $\log s$
  many times (since the total sparsity of the $g_i$'s reduces by at
  least a factor of half every time this step is executed), whereas
  the `fanin reduction' step can be performed at most $t-1$ times (as
  the fanin goes down by at least one for every such step). Finally,
  when this process of alternating steps ends, we have an equation of
  the form $g_i \cdot N_i = 0$ (Case $2$ of the fanin reduction
  step), where $g_i \neq 0$ and $N_i$ is a principal minor of a
  symbolic matrix $N$ of dimension at least $n - (\log s + t - 1)$
  such that at most $(\log s + t)$ entries of $N$ are set as
  constants. Now, if $\log s + t \le n - (\log s + t)$ the $N_i$ can
  never be zero (by Fact \ref{fac-imm-proj}) and hence we arrive at a 
  contradiction. Therefore, $s
  > 2^{n/2 - t}$. Now, the details of the sparsity reduction and
  the fanin reduction steps.

  Suppose, we have an equation $\sum_{i=1}^{\tau}{g_i N_i} = 0$ as
  mentioned above. Without loss of generality, assume that the minor
  $N_i$ is the minor of $N$ with respect to the $i^{th}$ diagonal
  element of $N$. Call all the variables $x_{ij}$ in $N$ with both
  $i,j> \tau$ as the \emph{white variables}. These are the variables
  that are present in every minor $N_i$ in the sum
  $\sum_{i=1}^{\tau}{g_i N_i}$. The variables $x_{ij}$ where both
  $i,j\leq \tau$ are called the \emph{black variables}, and the
  remaining are the \emph{grey variables}. By assumption, $c$ of the
  variables in $N$ are set as constants.

  \emph{Sparsity reduction step} - Say $x$ is a white variable that
  one of the $g_i$'s depends on. Writing each $g_i$ as a polynomial in
  $x$, there must be two distinct powers of $x$ amongst the $g_i$'s
  (for if not, then $x$ can be taken common across all $g_i$'s). Let
  $x^\ell$ be the lowest degree and $x^h$ be the highest. Dividing the
  entire equation $\sum_{i=1}^{\tau}{g_i N_i} = 0$ by $x^\ell$, we can
  further assume that $\ell = 0$. Each of the $g_i$'s and $N_i$'s can
  be expressed as, $g_i = g_{i,0} + x \cdot g_{i,1} + \cdots + x^h
  \cdot g_{i,h}$ and $N_i = N_{i,0} +x \cdot N_{i,1}$, where $g_{i,
    j}$'s and $N_{i,j}$'s are $x$-free.  Looking at the coefficients
  of $x^0$ and $x^{h+1}$ in the equation yields $ \sum_{i=1}^\tau
  g_{i,0}\cdot N_{i,0} = 0$ and $\sum_{i=1}^\tau g_{i,h} \cdot N_{i,1}
  = 0$.  Note that $N_{i,0}$'s can be thought of as principal minors
  of the $\eta \times \eta$ matrix $N'$ obtained by setting $x=0$ in
  $N$. And each of the $N_{i,1}$'s can be thought of as minors of the
  $(\eta - 1) \times (\eta - 1)$ matrix $N'$ which is the matrix
  associated with the minor of $N$ with respect to $x$. Since the
  monomials in $g_{i,0}$ and $x^h g_{i,h}$ are disjoint, either the
  total sparsity of the $g_{i,0}$'s or the total sparsity of the
  $g_{i,h}$'s is $\le$
  {\footnotesize$\mathcal{S}$}/2. Thus, one of the equations $
  \sum_{i=1}^\tau g_{i,0}\cdot N_{i,0} = 0$ or $\sum_{i=1}^\tau
  g_{i,h} \cdot N_{i,1} = 0$ yields an equation of the form
  $\sum_{i=1}^{\tau'}{g_i' N_i'} = 0$ with parameters $\tau',$
  {\footnotesize$\mathcal{S'}$}, $\eta'$, $c'$ as claimed before. (In
  case, we choose $\sum_{i=1}^\tau g_{i,h} \cdot N_{i,1} = 0$ as our
  next equation, we also set the variables in the same columns and
  rows of $x$ to constants in such a way that a $g_{i,h}$ stays
  nonzero. This is certainly possible over a characteristic zero
  field \cite{S80, Z79}.) The sparsity reduction step is performed
  whenever the starting equation $\sum_{i=1}^{\tau}{g_i N_i} = 0$ has
  a white variable among the $g_i$'s. When all the $g_i$'s are free of
  white variables, we perform the \emph{fanin reduction step}.

  \emph{Fanin reduction step} - When we perform this step, all the
  $g_i$'s consist of black and grey variables. Pick a row $R$ from $N$
  barring the first $\tau$ rows. Let $y_1,\cdots, y_\tau$ be the grey
  variables occuring in $R$ (these are, respectively, the variables in
  the first $\tau$ columns of $R$). Starting with $y_2$, divide the
  equation $\sum_{i=1}^{\tau}{g_i N_i} = 0$ by the largest power of
  $y_2$ common across all monomials in the $g_i$'s, and then set $y_2
  = 0$. This process lets us assume that there exists at least one
  $g_i$ which is not zero at $y_2 = 0$. On the residual equation,
  repeat the same process with $y_3$ and then with $y_4$ and so on
  till $y_\tau$. Thus, we can assume without loss of generality that
  in the equation $\sum_{i=1}^{\tau}{g_i N_i} = 0$ there is at least
  one $g_i$ that is not zero when $y_2, \ldots, y_\tau$ are set to
  zero. Observe that if $g_1$ is the only $g_i$ that stays nonzero
  under the projection $y_2 = \ldots = y_\tau = 0$ then $(g_1
  N_1)_{(y_2 = \ldots = y_\tau = 0)} = 0$, implying that $N_1 = 0$
  under the same projection - this is Case $2$ of the fanin reduction
  step mentioned earlier. Now, assume that there is a $g_i$ other than
  $g_1$ (say, $g_2$) that is nonzero under the projection $y_2 =
  \ldots = y_\tau = 0$. Set all the remaining variables of row $R$ to zero
  except $y_1$ - these are the white variables in $R$. Since the
  $g_i$'s are free of white variables (or else, we would have
  performed the `sparsity reduction' step), none of the $g_i$'s is
  effected by this projection. However, $N_1$ being a minor with
  respect to the first diagonal element of $N$, vanishes completely
  after the projection. Any other $N_i$ takes the form $y_1 \cdot
  N_i'$, where $N_i'$ is a principal minor of a $(\eta - 1) \times
  (\eta - 1)$ matrix $N'$ which is the matrix associated with the
  minor of $N$ with respect to $y_1$. Therefore, after the projection,
  the equation $\sum_{i=1}^{\tau}{g_i N_i} = 0$ becomes
  $\sum_{i=2}^{\tau} {\tilde{g_i} \cdot y_1 N_i' = 0} \Rightarrow
  \sum_{i=2}^{\tau} {\tilde{g_i} \cdot N_i' = 0}$, where $\tilde{g_i}$
  is the image of $g_i$ under the above mentioned projection and
  further $\tilde{g_2} \neq 0$. The $\tilde{g_i}$'s might still
  contain variables from the first column of $N$. So, as a final step,
  set these variables to values so that a nonzero $\tilde{g_i}$
  remains nonzero after this projection (the \cite{S80,Z79,DL78} lemma 
  asserts that such values exist in plenty). This gives
  us the desired form $\sum_{i=1}^{\tau'}{g_i' N_i'} = 0$ with
  parameters $\tau',$ {\footnotesize$\mathcal{S'}$}, $\eta'$, $c'$ as
  claimed before (Case 1 of the fanin reduction step mentioned
  earlier).
\end{proof}

\begin{replemma}{lem:det-ind-modulo}
  If $M_1,\cdots, M_t$ are distinct first order principal minors of
  $M$ and $\sum_{i=1}^{t}\alpha_i M_i$ $= 0\bmod \boldsymbol\ell_k$ (not all
  $\alpha_i = 0$) for independent linear polynomials
  $\boldsymbol\ell_k$, then $t + k \ge n$. 
\end{replemma}
\begin{proof}
  Assume that $t+k< n$ (with $t\ge 1$ it means $k\le n-2$). Since $\ell_1,\cdots,
  \ell_k$ are independent linear polynomials, the equation may be
  rewritten as $\sum_{i=1}^{t} \alpha_i M_i' = 0$ where $(M_i')$s are minors of
  the matrix $M'$ obtained by replacing $k$ entries of $M$ by linear
  polynomials in other variables. We shall call these entries as
  \emph{corrupted} entries.  Without loss of generality, we shall assume
  that $M_i'$ is the minor corresponding to the $i$-th diagonal
  variable and that all the $\alpha_i$'s are nonzero.
  \begin{claim}
    Each of the first $t$ rows and columns must have a corrupted entry. 
  \end{claim}
  \vskip -0.1in
  \begin{myproof}{Claim}
    Suppose the first row (without loss of generality) is free of any
    corrupted entry. Then, setting the entire row to zero would make all
    $M_i' = 0$ for $i\neq 1$. But since $\sum \alpha_i M_i' = 0$, this
    forces $M_1'$ to become zero under the projection as well. This
    leads to a contradiction as $M_1'$ is a determinant of an
    $(n-1)\times (n-1)$ symbolic matrix under a projection, and this
    can not be zero unless $k \geq n-1$ (by Fact \ref{fac-imm-proj}).
  \end{myproof}
  
  \noindent
  Since $n-k>t$, there must exist a set of $t-1$ rows
  $\inbrace{R_1,\cdots, R_{t-1}}$ of $M$ that are free of any corrupted entries.
  For each of these rows, set the $i$-th variable of row $R_i$ to $1$,
  and every other variable in $R_1,\cdots, R_{t-1}$ to zero. These
  projections make $M_i' = 0$ for all $i\neq t$ (as in these minors an entire row vanishes). 
  And since $\sum_{i=1}^{t}
  \alpha_i M_i' = 0$, this forces $M_t'$ to become zero under this
  projection as well. But under this projection, $M_t'$ just reduces
  (up to a sign) to the minor obtained from $M'$ by removing the
  columns $\inbrace{1,\cdots, t}$ and rows $\inbrace{R_1,\cdots,
    R_{t-1}}\union \inbrace{t}$. This is a determinant of an
  $(n-t)\times(n-t)$ symbolic matrix, containing at most
  $k-t$ corrupted entries, thus $k-t \geq n-t$ 
  (by Fact \ref{fac-imm-proj}). But then $k\ge n$, which contradicts our initial assumption.
\end{proof}

\subsection{Extensions to immanants}\label{sec:app_imm}

All the lower bound proofs use some very basic properties of
$\Det_n$. These properties are general enough that they apply to any
\emph{immanant}. For any character $\chi:S_n\to\C^{\times}$, recall the
definition of the immanant of an $n\times n$ matrix $M = (x_{ij})$:
$$
\text{Imm}_\chi(M) \quad=\quad \sum_{\sigma \in S_n}\chi(\sigma)
\prod_{i=1}^n x_{i,\sigma(i)}
$$
Since $\chi$ is a character, this in particular means that
$\chi(\sigma) \neq 0$ for any $\sigma\in S_n$

\begin{definition}[Immanant minor]
  The \emph{minor} of $\text{Imm}_\chi(M)$ with respect to the
  $(i,j)$-th entry is defined as
$$
\inparen{\text{Imm}_\chi(M)}_{i,j} \quad=\quad \sum_{\substack{\sigma
    \in S_n\\\sigma(i) = j}}\chi(\sigma) \prod_{k\neq i}
x_{k,\sigma(k)}
$$
\end{definition}
This may also be rewritten as a scalar multiple of
$\text{Imm}_{\chi'}(M_{ij})$ for a suitable character
$\chi':S_{n-1}\to \C^{\times}$, where $M_{ij}$ is the submatrix of $M$
after removing the $i$-th row and $j$-th column. From the definition, 
it follows directly that
the partial derivative of $\text{Imm}_\chi(M)$ with respect to $x_{ij}$ is
precisely the minor with respect to $(i,j)$. \\

The only crucial fact of determinants that is used in all the proofs
is that a symbolic $n\times n$ determinant cannot be zero when less
than $n$ of its entries are altered. 

\begin{fact}\label{fac-imm-proj}
  Let $M'$ be the matrix obtained by setting $c<n$ entries of $M$ to
  arbitrary polynomials in $\F[\mathbf{x}]$. Then for any character
  $\chi:S_n\to\C^\times$, we have $\text{Imm}_\chi(M') \neq 0$.
\end{fact}
\begin{proof}
  We shall say an entry of $M'$ is \emph{corrupted} if it is one of
  the $c$ entries of $M$ that has been replaced by a polynomial. 
  We shall prove this by carefully rearranging the rows and
  columns so that all the corrupted entries are above the
  diagonal. Then, since all entries below the diagonal are free, we
  may set all of them to zero and the immanant reduces to a single
  nonzero monomial. 
  
  Since less than $n$ entries of $M'$ have been altered, there exists
  a column that is free of any corrupted entries. By relabelling the
  columns if necessary, let the first column be free of any corrupted
  entry. Pick any row $R$ that contains a corruption and relabel the
  rows to make this the first row. This ensures that the first column
  is free of any corrupted entry, and the $(n-1)\times(n-1)$ matrix
  defined by rows and columns, $2$ through $n$, contain less than $c-1$
  corruptions. By induction, the $c-1$ corruptions may be moved above
  the diagonal by suitable row/column relabelling. And since the first
  column is untouched during the process, we now have all $c$
  corruptions above the diagonal. Now setting all entries below the
  diagonal to zeroes reduces the immanant to a single nonzero
  monomial.
\end{proof}

\noindent
With this fact, all our lower bound proofs of the determinant can be
rewritten for any immanant. 

\section{Conditional immanant lower bounds for depth-$D$ occur-$k$
  formulas}\label{sec:app_cond_lbnd}

In this section, we present a lower bound for depth-$D$ occur-$k$
formulas similar in spirit to
Theorem~\ref{thm:lowerbnd-depth4-occurk} by assuming the following conjecture
about determinant minors. 

\begin{conjecture}\label{conj:minor-conjecture}
  Let $M = (x_{ij})$ be an $n\times n$ matrix, and let $x_i$ denote
  the $i$-th diagonal variable $x_{ii}$. Let $M'$ be a projection of
  $M$ by setting $c=o(n)$ of the variables in $M$ to constants. Suppose the
  elements $x_1,\cdots, x_k$, where $k$ is a constant independent of
  $n$, are partitioned into non-empty sets $S_1,\cdots, S_t$. Consider
  $\mathcal{M}(\mathbf{S}_t)$, the set of $t^{\text{th}}$ order
  principal minors of $M'$, each by choosing a $t$-tuple $B\in
  S_1\times\cdots \times S_t$ as pivots. Over all possible choices of
  $B$, we get $m := |S_1|\cdots |S_t|$ many minors. Then for any set of
  diagonal variables $\mathbf{y}_m$ disjoint from $\mathbf{x}_k$, 
  $J_{\mathbf{y}_m}\inparen{\mathcal{M}(\mathbf{S}_t)}\ne0$.
\end{conjecture}

The conjecture roughly states that the different $t^{\text{th}}$-order
principal minors are algebraically independent. We will need a
generalization of Lemma~\ref{lem:diag-minor-sum} for the purposes of
this section.

\begin{lemma}\label{lem:diag-minor-sum-gen}
  Suppose $\{f_1,\cdots, f_s, g_1,\cdots, g_t\}$ are algebraically dependent 
  polynomials such that
  $\mathrm{trdeg}\inbrace{\mathbf{g}_t} = t$. Let $S\subseteq
  \mathbf{x}$ be a fixed set of variables of size at least $s +
  t$. Then there exists a set of $s+t$ variables
  $\mathbf{x}_{s+t}\subset \mathbf{x}$ and an equation of the form
  $$
  \sum_{i=1}^r c_i \cdot F_i\cdot G_i = 0\quad\quad \text{where } r\leq \binom{s+t}{t}
  $$
  such that each $c_i \in\F^*$, each $F_i$ is a distinct
  $s\times s$ minor of $\mathcal{J}_{\mathbf{x}_{s+t}\intersection
    S}(\mathbf{f}_s)$, each $G_i$ is a distinct $t\times t$ minor
  of $\mathcal{J}_{\mathbf{x}_{s+t}}(\mathbf{g}_t)$, and not all $G_i$'s
  are zero.
\end{lemma}

\noindent Note that we are not asserting the nonzeroness of $F_i$'s.
Also, Lemma~\ref{lem:diag-minor-sum} may be obtained from the
above lemma by taking $f_1 = \Det_n$, $s=1$ and $S$ to be the set of
diagonal variables. 

\begin{proof}
  The proof is along the lines of
  Lemma~\ref{lem:diag-minor-sum}. Amongst the nonzero $t \times t$
  minors of $\mathcal{J}_{\mathbf{x}}(\mathbf{g}_t)$, pick one (call
  the matrix associated with the minor, $N$) that maximizes the number
  of variables in $S$ indexing the columns of $N$. Without loss of
  generality, let $\mathbf{x}_t$ be the set of variables indexing the
  columns of $N$. Since $|S| \geq s + t$, there exists $s$ other
  variables in $S$, say $\inbrace{x_{1+t},\cdots, x_{s+t}}$.
  Consider the $(s+t)\times(s+t)$ minor of
  $\mathcal{J}_{\mathbf{x}}(\mathbf{f}_s \cup \mathbf{g}_t)$
  corresponding to the columns indexed by $\mathbf{x}_{s+t}$ - call
  the associated $(s+t) \times (s+t)$ matrix $\tilde{N}$.

  Since $\mathbf{f}_s,\mathbf{g}_t$ are algebraically dependent, we have that
  $\det(\tilde{N})=0$. Expanding $\det(\tilde{N})$ over all possible
  $s\times s$ minors in the first $s$ rows, we have an equation
  $$
  \sum_{U\subseteq \mathbf{x}_{s+t}, |U| = s} c_i \cdot F_U \cdot
  G_{U} =0
  $$where each $F_U$ is a distinct
  $s\times s$ minor of $\mathcal{J}_{\mathbf{x}_{s+t}}(\mathbf{f}_s)$,
  each $G_U$ is a distinct $t \times t$ minor of
  $\mathcal{J}_{\mathbf{x}_{s+t}}(\mathbf{g}_t)$, and $c_i \in\F^*$. 
  If $G_U$ is the minor with respect to variables
  $\mathbf{x}_{t}$, then $G_U = \det(N) \neq 0$ (by construction). It
  suffices to show that if $F_U$ is a minor indexed by variables
  outside $S$, then $G_U = 0$. This follows, just like in
  Lemma~\ref{lem:diag-minor-sum}, by the maximality assumption on
  choice of $\mathbf{x}_t$.
\end{proof}

\noindent The rest of this section shall be devoted to the proof of
the following theorem.

\begin{theorem}\label{thm:cond_lbnd_dDok}
  Assuming Conjecture~\ref{conj:minor-conjecture}, any depth-$D$
  occur-$k$ formula that computes $\Det_n$ must have size $s =
  2^{\Omega(n)}$ over any field of characteristic zero.
\end{theorem}

\noindent {\bf Proof idea:} The proof proceeds on the same lines as
Theorem~\ref{thm:lowerbnd-depth4-occurk}. If $T_1,\cdots, T_k$ is a
transcendence basis of gates at level $2$ computing the determinant,
then $\mathcal{J}_{\mathbf{x}}(\Det_n, T_1,\cdots, T_k)$ is a matrix
of rank $k$. This yields a non-trivial equation of the form $\sum
N_i^{(1)} \cdot G_i^{(1)} = 0$ where each of the $N_i^{(1)}$'s are
principal minors of $M = (x_{ij})$ and $G_i^{(1)}$'s are $k\times k$
minors of $\mathcal{J}_{\mathbf{x}}(T_1,\cdots, T_k)$. Here is where
we may use Lemma~\ref{lem:descent-jacobian} to remove common factors
and obtain an equation of the form $\sum N_i^{(1)} \cdot
\tilde{G_i}^{(1)} = 0$ where $\tilde{G_i}^{(1)}$ is a polynomial of
constantly many derivatives of polynomials computed at the next
level. The above equation may be thought of as a polynomial relation
amongst $\inbrace{N_i^{(1)}}\union
\inbrace{\text{Elem}(\tilde{G_i}^{(1)})}$. Applying
Lemma~\ref{lem:diag-minor-sum-gen} (with a suitable choice of $S_2$),
we get an equation of the form $\sum N_i^{(2)} \cdot G_i^{(2)} = 0$
where each $N_i^{(2)}$ is a minor of $\mathcal{J}_{S_2}\inparen{\inbrace{N_i^{(1)}} }$, and
$G_i^{(2)}$s are Jacobians minors of $\Union
\text{Elem}(\tilde{G_i}^{(1)})$. Again after removing common factors, this equation may be
intepreted as a polynomial relation amongst the entries of $N_i^{(2)}$
(which are minors of order $2$) and $\text{Elem}(\tilde{G_i}^{(2)})$. 

Repeating this argument, we finally reach the level of sparse
polynomials and obtain a non-trivial equation $\sum N_i^{(D-2)} \cdot
\tilde{G_i}^{(D-2)} = 0$, where each $N_i^{(D-2)}$ is a Jacobian minor of 
$(D-3)$-order minors, and each $\tilde{G_i}^{(D-2)}$ is a sparse polynomial. With a
slightly more careful choice of the sets $S_i$ in
Lemma~\ref{lem:diag-minor-sum-gen}, each of the minors $N_i^{(D-2)}$
would be a minor of $\mathcal{J}_{S_{D-2}}(\mathcal{M}(S_1,\cdots,
S_{D-3}))$. Assuming Conjecture~\ref{conj:minor-conjecture}, we can
show that such an equation is not possible unless the sparsity of the
$f_i$'s is \emph{large}, using a similar argument as in
Lemma~\ref{lem:sparse-comb-minors}.

\begin{lemma}
  Suppose $\Det_n$ is computed by a depth-$D$ occur-$k$ formula of
  size $s$. Then there exist variables $x_1,\cdots, x_R$ where $R =
  R(k,D)$, a partition of $\mathbf{x}_R$ into non-empty sets $S_1,\cdots, S_{D'}$, 
  ($D'\le(D-2)$)
  polynomials $f_1,\cdots, f_m$ (not all zero) where $m =
  |\mathcal{M}(\mathbf{S}_D)|^{O(R)}$ and each $f_i$ has sparsity at most
  $s^R$, such that
  $$
  \sum_{i=1}^m f_i \cdot N_i = 0
  $$
  where each $N_i$ is a minor of $\mathcal{J}_{\mathbf{x}}(\mathcal{M}(\mathbf{S}_{D'}))$
  indexed by diagonal variables.
\end{lemma}
\begin{proof}
  To begin with, $\Det_n = C(T_1,\cdots, T_m)$ where $T_1,\cdots,
  T_m$ are polynomials computed at the first level. So Lemma~\ref{lem:diag-minor-sum} gives 
  a starting equation, though we do not really have a sparsity bound on the $f_i$'s.
  The proof shall
  proceed by transforming this equation into another, involving lower
  level polynomials, till we get a sparsity bound.

  In general, we shall have an equation of the form
  $C_\ell(\mathcal{M}(S_1,\cdots, S_{\ell-1}), T_1^{(\ell)},\cdots,
  T_{r_\ell}^{(\ell)}) = 0$, where each $T_{i}^{(\ell)}$ is a
  derivative (of order at most $\ell$) of a polynomial computed at
  level $\ell$ of the circuit. Without loss of generality, we may
  assume that $\inbrace{T_1^{(\ell)},\cdots, T_{r_\ell}^{(\ell)} }$ are algebraically
  independent.  Let $m_\ell := |S_1|\cdots |S_{\ell-1}|$. Choose a set
  of diagonal elements $S_{\ell}$ of size $|\mathcal{M}(S_1,\cdots,
  S_{\ell-1})|+r_\ell$ that is disjoint from $S_1,\cdots,
  S_{\ell-1}$. Applying Lemma~\ref{lem:diag-minor-sum-gen} with
  $S_\ell$, we get an equation of the form
  $$
  \sum_i c_i^{(\ell)} N_i^{(\ell)}\cdot G_i^{(\ell)} = 0
  $$
  where $N_i^{(\ell)}$ is an $m_\ell\times m_{\ell}$ minor of
  $\mathcal{J}_{S_\ell}(\mathcal{M}(\mathbf{S}_{\ell-1}))$ indexed by diagonal
  variables, each $G_i^{(\ell)}$ is an $r_\ell\times r_{\ell}$ minor
  of $\mathcal{J}_{S_\ell}(\mathbf{T}_{r_{\ell}}^{(\ell)} )$. Since $C$ is an occur-$k$
  circuit, using an argument similar to
  Lemma~\ref{lem:descent-jacobian}, the above equation may be
  rewritten as
  $$
  V_\ell\cdot   \sum_i c_i^{(\ell)} N_i^{(\ell)}\cdot \tilde{G_i}^{(\ell)} = 0
  $$
  where each $\tilde{G_i}^{(\ell)}$'s is a polynomial function of at most
  $r_{\ell+1} := (\ell + 1)2^{\ell+1}\cdot k(r_\ell + m_\ell)r_\ell$ many
  derivatives of polynomials computed at level $\ell+1$. Note that
  $V_\ell$ cannot be zero as at least one $G_i^{(\ell)}$ was
  guaranteed to be nonzero by
  Lemma~\ref{lem:diag-minor-sum-gen}. Therefore, $\sum_i c_i^{(\ell)}
  N_i^{(\ell)}\cdot \tilde{G_i}^{(\ell)} = 0$. Since each $\tilde{G_i}^{(\ell)}$ is a
  polynomial function of $r_{\ell+1}$ derivatives at the next level, we
  now have $C_{\ell+1}(\mathcal{M}(S_1,\cdots, S_\ell),
  T_1^{(\ell+1)}, \cdots,$ $T_{r_{\ell+1}}^{(\ell+1)}) = 0$.
 
  Unfolding this recursion, we finally reach the level of sparse
  polynomials, at which point we have an equation of the form
  $$
  \sum_i c_i^{(D-2)} N_i^{(D-2)}\cdot \tilde{G_i}^{(D-2)} = 0
  $$
  and each $\tilde{G_i}^{(D-2)}$ is a $r_{D-2}\times r_{D-2}$ Jacobian minor of sparse
  polynomials. Hence, each $\tilde{G_i}^{(D-2)}$ is itself a polynomial of
  sparsity bounded by $s^{r_{D-2}}$ as claimed.
\end{proof}

We now have to show that an equation of the form $\sum f_i \cdot N_i = 0$ is
not possible unless one of the $f_i$'s has exponential sparsity. The above 
lemma guarantees that at least one of the $f_i$'s
are nonzero in this equation, but it could be the case that some of
the $N_i$'s are zero. This was not the case in the depth-$4$ lower
bound as each $N_i$ was just a determinant minor. However, in this
case they are jacobians of
minors. Conjecture~\ref{conj:minor-conjecture} asserts that the
$N_i$'s are nonzero, even if \emph{few} variables are set to
zero. This assumption would be enough to get the required lower bound.

\begin{lemma}
  Let $|\mathcal{M}(S_1,\cdots, S_D)| =: m$ be a constant and let
  $\inbrace{N_i}_{i\leq t}$ be
  distinct $m\times m$ minors of
  $\mathcal{J}_{\mathbf{x}}(\mathcal{M}(\mathbf{S}_D))$ where the columns of $N_i$ 
  are indexed by a
  set $T_i$ of diagonal variables of $M$ disjoint from $\Union_{j=1}^{D} S_j$. Suppose
  $f_1,\cdots, f_t$ are polynomials such that $\sum_{i=1}^{t} f_i \cdot N_i = 0$
  (not all $f_i$'s are zero). Then,
  assuming Conjecture~\ref{conj:minor-conjecture} is true, the total
  sparsity of the $f_i$'s is $2^{\Omega(n)}$.
\end{lemma}
\begin{proof}
  The proof is along the lines of the proof of
  Lemma~\ref{lem:sparse-comb-minors} and shall proceed by a similar
  series of \emph{sparsity reduction} and \emph{fanin reduction}
  steps to arrive at a contradiction. Throughout the proof,
  Conjecture~\ref{conj:minor-conjecture} shall assert that $N_i$'s
  stay nonzero (even when few variables are set to constants). We
  briefly describe the \emph{sparsity reduction} and the \emph{fanin
    reduction} steps and the rest of the proof would follow in
  essentially an identical fashion as the proof of
  Lemma~\ref{lem:sparse-comb-minors}.

  Without loss of generality, assume that $\inbrace{x_1,\cdots, x_r}$
  is the union of the sets $S_i$'s and $T_i$'s. Let $N$ refer to the
  matrix of indeterminates that the $N_i$'s are derived from. In our
  case, $N$ would be obtained by (possibly) setting few variables to constants in
  $M = (x_{ij})$. We'll refer to all the variables $x_{ij}$ where both
  $i,j> r$ as \emph{white} variables; these are present in every entry
  of each $N_i$. The variables $x_{ij}$ where both $i,j\leq r$ shall
  be called \emph{black} variables, and the rest
  called \emph{grey} variables. Here again, the \emph{sparsity
    reduction} step shall be applied whenever one of the $f_i$'s
  depends on a \emph{white} variable, otherwise the \emph{fanin reduction}
  steps shall be applied. \\

  \vskip -0.1in
  \emph{Sparsity-reduction} step - Suppose one of the $f_i$'s depend
  on a white variable $x$. Then each $N_i$ can be written as $N_i =
  N_{i,0} + \cdots + x^{m}N_{i,m}$, and $f_i = f_{i,0} + \cdots +
  f_{i,h}x^h$. One of the two equations corresponding to the
  coefficient of $x^0$ and $x^{h+m}$ yields a similar equation with
  sparsity reduced by a factor of $1/2$. Observe that $N_{i,0}$ is
  just $N_i\mid_{x=0}$, and hence the polynomials $\inbrace{N_{i,0}}$
  may be thought of as corresponding Jacobian minors of $N'$ obtained
  by setting $x=0$ in $N'$. Also, $N_{i,m}$ is obtained by replacing
  every entry of the matrix corresponding to $N_i$ by its minor with
  respect $x$. And hence, $N_{i,m}$ can be thought of as a
  corresponding Jacobian minor of $N_x$ obtained by taking the minor
  of $N$ with respect to $x$. Thus the two equations corresponding to
  the coefficient of $x^0$ and $x^{h + m}$ are indeed of the same
  form as $\sum f_i N_i = 0$. (In the case of the coefficient of
  $x^{h+m}$, we need to set other variables in the row/column
  containing $x$ as in the proof of
  Lemma~\ref{lem:sparse-comb-minors})\\

  \vskip -0.1in
  \emph{Fanin reduction} step - Without loss of generality, let 
  $x_1\in T_1\setminus T_2$. Pick a row $R$ of
  $N$ barring the first $r$ rows, and let $y_1,\cdots, y_r$ be the
  grey variables in $R$ (where $y_1$ is in the same column as
  $x_1$). By a similar process as in the proof of
  Lemma~\ref{lem:sparse-comb-minors}, we can assume that at least one
  $f_i$ is nonzero when $y_2,\cdots, y_r$ are set to zero. 

  If one of the $f_i$'s become zero when $y_2,\cdots,y_r = 0$, then
  pick any white variable $y$ in row $R$ and set every variable in row
  $R$ to zero besides $y$. This would ensure that the fanin of the
  equation reduces and each $N_i$ is now $y^{m}\cdot N_i'$. Each
  $N_i'$ may be thought of as being obtained from $N_y$, the minor of
  $N$ with respect to $y$. The other variables in the column of $y$
  can be set to values to ensure that the $f_i$'s stay nonzero to
  obtain an equation of the form $\sum f_i' N_i' = 0$ of reduced
  fanin.

  If none of the $f_i$'s become zero when $y_2,\cdots, y_r = 0$, then
  set every variable in row $R$ other than $y_1$ to zero. This ensures
  an entire column of the matrix corresponding to $N_1$ becomes zero
  (as $x_1$ indexes one of the columns of $N_1$), and hence $N_1$
  becomes zero. On the other hand, $N_2$ remains nonzero and each
  surviving $N_i$ can be written as $y_1^{m}\cdot N_i'$, where
  $N_i'$ is the corresponding Jacobian minor of $N_{y_1}$. Again, the
  other variables in the column of $y_1$ can be set to values to
  ensure that $f_i$'s stay nonzero and we obtain an
  equation $\sum f_i' N_i' = 0$ of reduced fanin.   \\

  \vskip -0.1in
  As in the proof of Lemma~\ref{lem:sparse-comb-minors}, we eventually
  obtain an equation of the form $f_1 N_1 = 0$ where $f_1 \neq 0$ thus
  implying that $N_1 = 0$. The number of variables that have been set
  to constants is bounded by $t + \log S$ where $S$ is the initial
  total sparsity of the $f_i$'s, and $N_1$ is a Jacobian minor of a
  symbolic matrix of dimension $n - (\log S + t
  -1)$. Conjecture~\ref{conj:minor-conjecture} asserts that $N_1$
  would be nonzero unless $\log S + t = \Omega(n - (\log S + t -1))$,
  or $S = 2^{\Omega(n)}$. 
\end{proof}

\noindent That concludes the proof of Theorem~\ref{thm:cond_lbnd_dDok} as well.

\end{document}